\documentclass[12pt]{iopart}
\usepackage{graphicx}
\usepackage{caption}
\usepackage{subcaption}
\usepackage{titlesec}
\usepackage[width=\textwidth,font=small,labelfont=bf]{caption}
\usepackage{hyperref}
\usepackage{amsmath}
\usepackage[ backend=biber, sorting=none, style=nature ]{biblatex}

\DeclareFieldFormat{url}{\mkbibacro{URL}\addcolon\space\url{#1}}

\DeclareFieldFormat{eprint:arxiv}{%
  \mkbibacro{URL}\addcolon\space\href{https://arxiv.org/abs/#1}{https://arxiv.org/abs/#1}%
}

\DeclareBibliographyDriver{incollection}{%
  \usebibmacro{bibindex}%
  \usebibmacro{begentry}%
  \usebibmacro{author/editor+others/translator+others}%
  \setunit{\labelnamepunct}\newblock
  \usebibmacro{title}
  \newunit\newblock
  \printtext{in\space}%
  \usebibmacro{maintitle+booktitle}%
  \newunit\newblock
  \usebibmacro{byeditor+others}%
  \newunit\newblock
  \printfield{pages}%
  \newunit\newblock
  \printfield{publisher}%
  \setunit*{\addcomma\space}%
  \printlist{location}%
  \setunit*{\addcomma\space}%
  \printfield{year}%
  \newunit\newblock
  \usebibmacro{doi+eprint+url}%
  \usebibmacro{finentry}%
}

\hypersetup{
    colorlinks=true,
    urlcolor=blue, citecolor = blue, linkcolor = blue}

\begin{document}

\newcommand{\Fstatistic}{$\mathcal{F}$-statistic}
\newcommand{\fstatistic}{$2\!\mathcal{F}$}
\newcommand{\Tobs}{T_{\rm obs}}
\newcommand{\vs}{{\it vs.}}

\title[Monitoring CW Hardware Injections in the LIGO O4 run]{Monitoring of Continuous-Wave Hardware Injections in LIGO Interferometers during the O4 Observing Run}

\author{Preet Baxi$^1$, Jessica Leviton$^{1*}$, Eilam Morag$^{2\dagger}$,
  Matthew Pitkin$^{3,4}$ and Keith Riles$^1$}

\address{$^1$ Dept. of Physics, University of Michigan, Ann Arbor, U.S.A.}
\address{$^2$ Dept. of Electrical and Computer Engineering, University of Michigan, Ann Arbor, U.S.A.}
\address{$^3$ CEDAR Audio Ltd., Cambridge, United Kingdom}
\address{$^4$ School of Physics \& Astronomy, University of Glasgow, Glasgow, United Kingdom}
\address{$*$ Now at Aysling Corporation, Ann Arbor, U.S.A.}
\address{$\dagger$ Now at Dept. of Electrical \&\ Computer Engineering, University of California, Los Angeles, U.S.A.}
\ead{preetb@umich.edu, kriles@umich.edu}

\vspace{10pt}
\begin{indented}
\item[]\today
\end{indented}

\begin{abstract}
Although there have now been hundreds of transient gravitational-wave detections of merging compact stars by the LIGO-Virgo-KAGRA (LVK) detector network, no continuous-wave (CW) signals have yet been discovered. To ensure that such signals, expected to be exceedingly weak, can be detected in the ongoing O4 observing run by coherent integration over years, simulated waveforms (``hardware injections'') are injected directly into the LIGO data by continuously modulating the positions of the interferometer mirrors so as to mimic nearly sinusoidal signals from fast-spinning galactic neutron stars. A set of 18 such simulated CW sources are injected with signal frequencies spanning much of the LIGO detection band and with varying sky locations. By verifying the successful recovery of the simulated signals, including preservation of absolute phase over as many as 10$^{11}$ signal cycles, we validate our understanding of detector response and end-to-end search pipelines, including data cleaning. Daily and weekly monitoring of the signal reconstruction is meant to catch any unexpected sudden changes in interferometer response, to verify that signal-to-noise ratio increases as expected and to verify that simulated source parameters are recovered correctly. We describe three methods of monitoring: 1) a highly templated matched filter to extract signal amplitude and phase precisely; 2) a frequentist \Fstatistic\ evaluation that marginalizes over amplitude, phase and orientation of the star; and 3) a Bayesian reconstruction of the source parameters together with noise characterization. Results from each method are shown, with emphasis on the new templated method, which yields precise measurement of the critical phase offset parameter and therefore validates understanding of absolute timing delays in the detector response and data stream.
\end{abstract}


\section{Introduction}

Since the first detection of gravitational waves from merging black holes in September 2015~\cite{Abbott_2016} by the LIGO detector, there have been 90 published detections of compact star mergers by the LIGO-Virgo-KAGRA (LVK) Collaboration from the first three observing runs (O1--O3)~\cite{GWTC-3}. In addition, there were nearly 300 public alerts~\cite{publicalerts} of LVK merger detections in low latency by real-time analyses during the recently completed O4 run, which began May 24, 2023. Despite this success in detection of transient signals, no persistent gravitational-wave signals, continuous-wave (CW) or stochastic, have been detected.

\vspace{8pt}

\noindent CW gravitational signals are expected to be emitted by rapidly spinning neutron stars with a small non-axisymmetry (mass or mass-current quadrupole asymmetry), but with signal amplitudes at the Earth several orders of magnitude weaker than the amplitudes of transient signals detected to date. Detecting such weak signals requires integration over long time intervals, with the best possible sensitivity accessible via fully coherent integrations over full observing spans~\cite{Riles_2023}.

\vspace{8pt}

\noindent Reliably extracting such signals requires an accurate understanding of detector response over the observing time, including absolute phase response over as many as 100 billion signal cycles. To help validate that understanding, the LIGO Scientific Collaboration uses a ``hardware injection'' system~\cite{Biwer_2017} that allows direct simulation of gravitational-wave signals via modulation of the positions of the interferometer mirrors\footnote{The Virgo and KAGRA
Collaborations rely on unmodulated actuations for similar validation}.

\vspace{8pt}

\noindent This direct modulation is carried out via a ``photon calibrator'' consisting of an auxiliary laser directed at one of the interferometer mirrors at a known incident angle and known intensity, which is modulated quasi-sinusoidally~\cite{photoncalibration}. The actuation response function is relatively stable (approximately that of a free mass for frequencies well above the mirror's pendulum frequency) once the absolute scale factor and phase dependence are determined through meticulous pre-run calibration. This stability makes the actuation nearly independent of the bulk of rest of the LIGO calibration process, which takes into account frequency-dependent and time-dependent effects of the detector response function. 

\vspace{8pt}

\noindent Hence the monitoring of CW injections, as measured in the gravitational-wave data channel constructed via that detector response, provides a key validation of the calibration. In addition, because additional cleaning of the data channel for both transient instrumental glitches and instrumental spectral lines is often carried out to improve sensitivity to astrophysical signals, the hardware injection recovery can verify the cleaning has done no harm, and where expected, has indeed improved sensitivity.

\vspace{8pt}

\noindent A secondary benefit of the hardware injections (but not one that by itself justifies the injection campaign) is allowing validation of search programs, one that is independent of the varying methods applied by different CW search programs for efficient large-scale ``software injections'' used in tuning algorithmic parameter choices and in determining signal sensitivities~\cite{Abbott_O4a_allsky,Tau_2025_binary}.

\vspace{8pt}

\noindent This paper describes three different monitoring programs used both for ``real time'' monitoring (cadence: once per day) to catch potential unintended detector response changes and for longer-term analysis and validation of data cleaning techniques. The first method uses exact templates for the intended injections, allowing only amplitude and phase offsets to vary. The second method uses the venerable \Fstatistic\ detection statistic, which serves as the foundation of a multitude of CW searches for different potential sources~\cite{Riles_2023} and allows a frequentist estimate of amplitude, phase offset and two angles describing the orientation of the simulated star. The third method applies a Bayesian statistic used in searches for CW radiation from known pulsars~\cite{Dupuis_2005}. Collectively, these programs have helped validate both the nominal calibration of the gravitational strain channel and cleaning techniques used to remove deterministic instrumental contaminations during the O4 observing run,

\subsection{Continuous Gravitational Waves}

Continuous gravitational waves represent a distinct class of gravitational wave signals~\cite{Riles_2023}. This radiation, expected from rapidly rotating galactic neutron stars, arises from slight stellar asymmetry in mass quadrupole moment or mass current quadrupole moment~\cite{Zimmermann_1979,rmode1,rmodes2,rmodes3,rmodes4,rmodes5,Jones_2002,Melatos_2005,Owen_2005,Glampedakis_2018}. Those asymmetries lead to much weaker signal amplitudes, however, than what has been observed from compact binary mergers, despite the much larger distances of the merger sources. Long integrations (months to years) are expected to be necessary to detect CW sources, with the best sensitivity requiring phase agreement between the actual signal and the model used to extract the signal.

\vspace{8pt}

\noindent For a mass quadrupole moment asymmetry, the amplitude $h_0$ of the continuous wave generated by a spinning neutron star is related to its equatorial ellipticity $\epsilon$ and its distance $d$ from Earth. The equation for the amplitude can be derived to be:~\cite{Riles_2023}:

\begin{equation}
    h_0 = \frac{16\pi^2 G \epsilon I_{zz} f^2}{c^4 d},
\end{equation}

\noindent where $G$ is the gravitational constant, $c$ is the speed of light, $I_{zz}$ is the moment of inertia of the neutron star about its rotational axis $z$, and $f$ is the rotational frequency of the neutron star.

\vspace{8pt}

\noindent The equatorial ellipticity $\epsilon$, is defined as:

\begin{equation}
    \epsilon = \frac{|I_{xx} - I_{yy}|}{I_{zz}}
\end{equation}

\noindent where $I_{xx}$ are $I_{yy}$ are the other two moments of inertia along the principal axes of the neutron star.

\vspace{8pt}

\noindent Based on the absence of detections from previous searches and consistent with astrophysical knowledge, we expect plausible long-lived continuous gravitational waves to have a strain less than $10^{-24}$ and to originate from galactic sources, at distances of O(kpc)\cite{Abbott_2019_CWO2,Abbott2022_O3_CWsearch}. It is only the long-lived nature of these waves that offers hope for their detection as we can integrate data over long durations, to produce a statistically significant Signal-to-Noise Ratio (SNR). For a coherent integration and stationary detector noise, the SNR is expected to increase proportionally to the square root of the observation time as:

\begin{equation}
\text{SNR} \quad \propto \quad {h_0 \sqrt\frac{\Tobs}{S(f)}}
\end{equation}

\noindent where \( h_0 \) is the strain amplitude, $\Tobs$ is the total observation time, and $S(f)$ represents the power spectral noise density~\cite{Brady_1998,Jaranowski_1998} at the frequency $f$ of the signal. As energy and angular momentum are lost by the star to gravitational waves, the spin frequency and the CW signal frequency (double the spin frequency) should decrease slowly, leading to a small negative first derivative $df/dt$. In the LIGO injections described in the next section, only the first derivative is simulated, for simplicity, and signal amplitude is held constant within a given observing run, although a true CW source would have higher-order frequency derivatives that become important over long integration times and would have slight amplitude reduction as the spin frequency declines~\cite{Riles_2023}.

\vspace{8pt}

\noindent The eventual detection of continuous gravitational waves from a galactic neutron star should yield insight into the neutron star equation of state, and in the event of successful identification of electromagnetic radiation from the same star, should provide a splendid multimessenger source, one that might well be studied for years and decades to come, to understand both the dynamics of these exotic objects and to probe fundamental physics in a regime inaccessible to laboratory experiments~\cite{Lasky_2015}.

\subsection{LIGO and its Hardware Injection Monitoring: an overview}

Hardware injections~\cite{Biwer_2017} make up one element of the LIGO detector’s~\cite{Aasi_2015,Ganapathy_2023,Jia_2024,Soni_2025,Capote_2025} calibration and validation framework~\cite{Abbott_2017,Sun_2020}, confirming our ability to identify and reconstruct gravitational wave signals. These injections use physical modulations of the detector's test mass mirrors to simulate the effects of a gravitational wave, introducing controlled signals into the interferometric data stream. The objectives of hardware injections are to verify the fidelity of signal reconstruction pipelines, ensure the accuracy of amplitude and phase calibrations, and evaluate the robustness of data analysis methods under varying signal conditions~\cite{Biwer_2017}.

\vspace{8pt}

\noindent Continuous wave injections pose a technical challenge because of their extended duration~\cite{Riles_2023} and the need to maintain absolute phase coherence. To address this challenge, the LIGO injection system uses the same barycentering software~\cite{LALSuite_2018} used in published CW searches and in software injections, along with a data buffering mechanism~\cite{Biwer_2017} to ensure smooth, continuous insertion of signal waveforms without unintended discontinuities in injected waveforms. Because the CW injections are extremely narrowband, it has proven convenient to use a frequency-domain technique in which a Fourier transform of injected strain is computed for short intervals of data with a frequency-domain inverse actuation response function applied to the transform before inverting to create a time-domain waveform for injection into the photon calibrator system. The inverse actuation function includes a correction for time delays in the injection system itself. Although the software infrastructure exists for direct time-domain injection of CW signals~\cite{Biwer_2017}, it was found in the O1 observing run that such injection of CW signals leads to high-frequency noise injection.

\vspace{8pt}

\noindent Daily reconstructions of hardware injections, based on the previous day's data and based on the cumulative data, are carried out to ensure there have been no significant unintended changes in detector response. This check is most important each week, following regular Tuesday maintenance periods. Graphs are created and inspected to track trends in amplitude recovery, phase stability, and the expected growth in SNR over time. Additionally, cumulative analyses over the observation period to date allow tracking of detector stability, aiding in identifying drifts, systematic biases or spectral line contaminations that can affect sensitivity~\cite{Covas_PRD_2018}.

\vspace{8pt}

\noindent In CW searches, the hardware injections provide a set of sources for validating analysis programs. Because most CW search programs are designed to analyze data stretches of months to years in duration using narrowband spectral bands at a time, it is typically not feasible to carry out software injection campaigns by superposing time-domain strain waveforms on the data and then computing Fourier transforms. Instead, more computationally efficient techniques, tailored to each search program, are used in which Fourier coefficients or derived quantities are superposed. The associated variation in the technical implementations of injections used for tuning and validating different pipelines leaves open the possibility of inconsistency across different programs. The existence of a set of hardware injections already present in the data means that performance can be evaluated and compared without ambiguity across different programs. Validation of results with hardware injections is a routine requirement in LIGO-Virgo-KAGRA CW analyses.

\vspace{8pt}

\noindent As noted above, however, despite this utility, one would not implement hardware injections for this purpose alone. One could, in principle, simply create a data set copy with software injections to allow unambiguous cross-pipeline comparisons. The primary benefit and justification for hardware injections is to validate assumptions about detector response, including the effects of calibration and data cleaning, especially the self-gating of large transients~\cite{zweizig_2021,davisetal}, to provide a true end-to-end validation of the search.

\section{LIGO CW Hardware Injections}

\noindent The CW injections are meant to simulate neutron stars in both isolated and binary systems~\cite{LALSuite_2018}. These signals span a range of frequencies and amplitudes, ensuring coverage of the detector’s operational bandwidth. Key parameters such as nominal source frequency (at a particular reference time), strain amplitude, phase, and polarization are defined, enabling comparisons between the intended injected signals and the corresponding reconstructed values~\cite{Biwer_2017}. Below we describe the details of the signal model and the parameter choices made for the 18 different stars simulated during the O4 observing run.

\subsection{Signal Model for Continuous Wave Detection}

Carrying out CW searches in LIGO data with optimum sensitivity requires correcting for signal frequency and amplitude modulations. The frequency (or phase) modulations result from the detector's motion with respect to an approximately inertial frame of reference (taken to be the solar system barycenter). For a star in a binary system, there are also frequency modulations from the orbital motion of the source. Amplitude modulations arise from the changing orientations of the LIGO interferometers with respect to the line of sight to the star. The relative frequency modulations due to the Earth's daily rotation are O(10$^{-6}$), and those due to the Earth annual orbit around the solar system barycenter (SSB) are O(10$^{-4}$). The amplitude modulations are O(1) and have periodicities of 12 hours and 24 hours (sidereal)~\cite{Jaranowski_1998}. Here we provide more detail on some of these modulations in the form of the signal model. The barycentering code used to account for these modulations is based on the same model used in the pulsar astronomy community~\cite{Hobbs_2006_TEMPO2} and includes General Relativistic effects.

\vspace{8pt}

\noindent We treat a continuous gravitational wave emitter as an isolated, spinning, rigid, triaxial ellipsoid. The strain waveform perceived by an interferometer can be modeled as:

\begin{equation}
    h(t) = F_+ (t; \psi) h_+ \cos \Phi(t) + F_\times (t; \psi) h_\times \sin \Phi(t),
\end{equation}

\noindent where $h_+$ and $h_\times$ are the amplitudes of two orthogonal, linearly polarized signals defined with respect to the spin axis of the star:

\begin{equation}
  \label{eqn:polarizations}
    h_+ = h_0 \frac{1 + \cos^2 \iota}{2}; \qquad {\rm and} \qquad h_\times =  h_0 \cos \iota,
\end{equation}

\noindent where \( h_0 \) is the amplitude of the gravitational wave, \( \psi \) denotes the polarization angle, \( \iota \) is the inclination angle of the neutron star’s spin axis relative to the line of sight, and \( \Phi(t) \) describes the time-dependent phase of the signal~\cite{Jaranowski_1998}. The detector response functions, \( F_+ (t; \psi) \) and \( F_\times (t; \psi) \), capture the directional sensitivity of the interferometer to the plus and cross polarizations of the gravitational wave~\cite{Jaranowski_1998}.

\vspace{8pt}

\noindent The phase evolution, \( \Phi(t) \), plays a pivotal role in accurate signal modeling. It can be expressed as:

\begin{equation}\label{eq:phase_evolution}
  \Phi(t) = \Phi_0 + 2\pi \sum_{m=0}^N \frac{f^{(m)}(\tau_{\text{ref}}) \, \tau(t)^{m+1}}{(m+1)!}
\end{equation}

\vspace{8pt}

\noindent where \( \tau(t) \) is the SSB arrival time of the signal, which includes relativistic effects such as Shapiro delay and Einstein delay. The term \( \tau_{\text{ref}} \) represents a reference time, at which the frequency \( f \) and its higher derivatives \( f^{(m)} \) are defined. The precision of this phase model is crucial for retaining coherence over long observation periods.

\vspace{8pt}

\noindent Modulations arise on daily and annual timescales due to the Earth’s rotation and orbital motion around the Sun. These effects can be described as:

\begin{equation}
    \tau(t) = t + \frac{\mathbf{r}(t) \cdot \hat{\mathbf{n}}}{c} - \tau_{\text{ref}}
\end{equation}

\vspace{8pt}

\noindent where \( \mathbf{r}(t) \) is the position vector of the detector relative to the SSB, \( \hat{\mathbf{n}} \) is the unit vector pointing from the SSB toward the source, and \( c \) is the speed of light.

\subsection{Parameters of O4 Injections}

A campaign of 18 CW hardware injections (indexed from 0 to 17) has been carried out during the O4 run, comprising 16 isolated and 2 binary systems. These injections have the same source parameters as those used in previous observing runs~\cite{Abbott2022_O3_CWsearch}, except for amplitudes which have been reduced in keeping with frequency-dependent reductions in detector noise levels. Six of the injections spread across
the band are given a high enough amplitude to be detected in coherent integration of a single day of data.
These six loud injections are used for verifying on a daily basis that no unintended large changes have
occurred in the calibration, such as might occur from the addition of an inadvertent time delay or
even sign flip in the digitally controlled electronics.    

\vspace{8pt}

\noindent The sources selected for the CW O4 hardware injections span a frequency range of 12 to 2991 Hz. The O4 campaign includes 6 loud simulations (injections 1, 4, 5, 6, 10 and 14), {\it i.e.}, having relatively high SNRs for daily monitoring and 12 weaker injections for longer-term monitoring at other nominal frequencies. Table~\ref{tab:parameters1} lists the parameters of all 18 injections. Additional orbital parameters for the two binary simulations (injections 16 and 17) are listed in Table~\ref{tab:parameters2}.

\begin{table}
\begin{center}
  \noindent \scalebox{0.85}{%
    \begin{tabular}{|rrcccccccc|}\hline
      \strut Inj & $f_0$  & $\dot f$ & $A_+$ & $A_\times$ & $h_0$ & $\cos(\iota)$ & $\psi$ & Dec   & RA  
      \\
      \strut        & (Hz)      & (Hz/s)      &         &               &     &                   & (rad)   & (rad) & (rad) 
      \\ \hline
      0 &  265.57710520$^1$ & -4.15e-12 & 3.27e-26 & 3.19e-26 & 4.01e-26 &  0.795 &  0.770 & -0.981 &  1.249 \\ 
      1 &  849.08329620$^1$ & -3.00e-10 & 1.69e-25 & 1.29e-25 & 2.79e-25 &  0.464 &  0.356 & -0.514 &  0.653 \\ 
      2 &  575.16357300$^1$ & -1.37e-13 & 2.98e-26 & -2.97e-26 & 3.20e-26 & -0.929 & -0.222 &  0.060 &  3.757 \\ 
      3 &  108.85715940$^1$ & -1.46e-17 & 6.53e-26 & -1.05e-26 & 1.30e-25 & -0.081 &  0.444 & -0.584 &  3.113 \\ 
      4 & 1403.16333100$^1$ & -2.54e-08 & 2.64e-25 & 1.36e-25 & 4.90e-25 &  0.277 & -0.648 & -0.218 &  4.887 \\ 
      5 &   52.80832436$^1$ & -4.03e-18 & 2.42e-25 & 1.85e-25 & 3.99e-25 &  0.463 & -0.364 & -1.463 &  5.282 \\ 
      6 &  148.71902570$^1$ & -6.73e-09 & 1.64e-25 & -4.92e-26 & 3.20e-25 & -0.154 &  0.471 & -1.142 &  6.261 \\ 
      7 & 1220.97958100$^1$ & -1.12e-09 & 5.63e-26 & 5.42e-26 & 7.16e-26 &  0.757 &  0.512 & -0.357 &  3.900 \\ 
      8 &  194.30831850$^1$ & -8.65e-09 & 4.17e-26 & 6.12e-27 & 8.28e-26 &  0.074 &  0.170 & -0.583 &  6.133 \\ 
      9 &  763.84731650$^1$ & -1.45e-17 & 4.75e-26 & -4.25e-26 & 6.87e-26 & -0.619 & -0.009 &  1.321 &  3.471 \\ 
      10 &  26.35891290$^2$ & -8.50e-11 & 6.19e-25 & -6.19e-25 & 6.26e-25 & -0.988 &  0.615 &  0.748 &  3.867 \\ 
      11 &  31.42485980$^2$ & -5.07e-13 & 1.76e-25 & -1.04e-25 & 3.17e-25 & -0.329 &  0.412 & -1.017 &  4.976 \\ 
      12 &  39.72760970$^2$ & -6.25e-09 & 1.32e-25 & 1.16e-26 & 2.63e-25 &  0.044 & -0.068 & -0.296 &  5.792 \\ 
      13 &  12.43000000$^2$ & -1.00e-11 & 1.32e-24 & 0.00e+00 & 2.65e-24 &  0.000 &  0.000 &  0.250 &  0.250 \\ 
      14 & 1991.0926000$^2$ & -1.00e-12 & 3.80e-25 & 0.00e+00 & 7.61e-25 &  0.000 &  1.000 & -0.250 &  5.250 \\ 
      15 & 2991.0926000$^2$ & -1.00e-12 & 2.04e-25 & 0.00e+00 & 4.09e-25 &  0.000 &  1.000 & -0.250 &  5.250 \\ 
      16 &  234.5670000$^3$ & 0.00e+00 & 1.21e-25 & 1.15e-25 & 1.59e-25 &  0.725 &  4.084 & -0.273 &  0.349 \\ 
      17 &  890.1230000$^3$ & 0.00e+00 & 6.37e-26 & 6.04e-26 & 8.39e-26 &  0.720 &  4.084 & -0.273 &  1.919 \\ 
      \hline
  \end{tabular}}%
  \caption{Parameters describing the 18 CW gravitational wave injections. The \textit{Inj} column lists the injection index, numbered sequentially.
    $f_0$ represents the signal frequency at a certain reference time.
    The spin-down rate, $\dot f$ (Hz/s), gives the (fixed) first derivative of the signal frequency.
    $A_+$ and $A_\times$ denote the amplitudes of the gravitational wave signals in the two orthogonal polarization states, $+$ and $\times$, respectively,
    while $h_0$ provides the overall strain amplitude of the gravitational wave, combining the contributions from both polarizations.
    The parameter $\cos(\iota)$ represents the cosine of the inclination angle $\iota$, which describes the orientation of the star's
    spin axis relative to the line of sight to the observer.
    The parameter $\psi$ specifies the angle (radians) of the projected spin axis on the plane of the sky.
    The \textit{Dec} and \textit{RA} columns give the simulated star’s declination and right ascension in radians.
    Orbital parameters for the binary-system injections 16-17 are listed in Table~\ref{tab:parameters2}.\\
  {\footnotesize $^1$Reference time: GPS \>\>751680013.}\\
  {\footnotesize $^2$Reference time: GPS \>\>930582085.}\\
  {\footnotesize $^3$Reference time: GPS 1230336018.}}
  \label{tab:parameters1}
\end{center}
\end{table}

\begin{table}
\begin{center}
  \noindent \scalebox{0.85}{%
    \begin{tabular}{|rccccc|}\hline
     \strut Inj & $T_{\rm orb}$  & $a\sin(i)$   & $T_{\rm peri}$ & $\Phi_{\rm peri}$ & Ecc. 
     \\
     \strut     & (s)           & (lt$\cdot$s) &   (s)        &  (rad)           &     
     \\ \hline 
     16         &  83941.2      & 2.35         &  1230336018  &  1.23456         & 0.0 \\
     17         &  83941.2      & 2.35         &  1230379218  &  2.13456         & 0.0 \\
      \hline
  \end{tabular}}%
  \caption{Orbital parameters describing the two binary-system CW gravitational wave injections 16 and 17.
    $T_{\rm orb}$ gives the period of the orbit.
    The projected semi-major axis is $a\sin(i)$ is given in light-seconds.
    The GPS time of periapsis is $T_{\rm peri}$. 
    The argument of periapsis is given by $\Phi_{\rm peri}$ (radians).
    The eccentricities are both zero, however, for these circular orbits.
    These orbital parameters are chosen to be similar enough to those expected for
    the low-mass X-ray binary system Scorpius X-1 to make them useful for pipeline validation
    while not so close as to contaminate actual searches for Scorpius X-1.}
  \label{tab:parameters2}
\end{center}
\end{table}

\section{Methodology}

\noindent The methods used to monitor hardware injection analysis are presented in the following subsections: a template generation method, a \Fstatistic\ approach, and a Bayesian inference. Each method has specific strengths, to be discussed below.

\subsection{Template Generation Method}

The template generation method of injection recovery exploits the expected phase and amplitude modulations of
the signal and a matched filter approach to extract estimates of true injection amplitude and phase offset. Although all three monitoring methods presented in this article include some degree of templating, since all assume the correct sky location and at least approximately correct frequency evolution, the template generation method does so to the greatest degree. 

\def\dki{d^k[i]}
\def\Dki{D^k[i]}
\def\ski{s_{\rm T}^k[i]}
\def\Ski{S_{\rm T}^k[i]}

\vspace{8pt}

\noindent The observation span is split into segments of duration
$T_{\mathrm{SFT}}$ = 1800 s for which discrete Fourier transforms are computed, known as ``short Fourier transforms'' or SFTs. The SFTs for the data for each detector, used for spectral monitoring and astrophysical figures of merit, are already available at the start of each UTC day's injection monitoring run. That run begins several hours after the corresponding day's end, to allow for data transfer delays from the observatories to the Caltech computing cluster, where monitoring is run centrally. Parallel sets of narrowband SFTs for each day and for each injection are then computed for a pure signal template identical to what is supposed to have been injected in the data, using LIGO Analysis Library software~\cite{LALSuite_2018},
including a python-code wrapper~\cite{Wette2020}. The band of each SFT for a given injection is centered on the injected signal's nominal detector-frame frequency at the start of that SFT's time interval.

\vspace{8pt}

\noindent In the following, denote the $k$th data set SFT as $d^k$ and the complex Fourier coefficient contained in its $i$th bin as $\dki$, with the corresponding signal template quantities $s_{\rm T}^k$ and $\ski$. The (real) powers in these bins are simply the absolute squares of these values: $\Dki = |\dki|^2$ and $\Ski = |\ski|^2$.

\vspace{8pt}

\noindent The analysis for each SFT $k$ is performed in two steps. First, a discrete frequency band (selection window) is determined that captures at least 99\% of the total signal template power. Second, an inner product of the data and the complex conjugate of the signal template bins within the selection window is computed, from which estimates of signal amplitude and phase offset can be computed, both for SFT $k$ alone and for the cumulative set of SFTs from 1 to $k$ in the observing run. Uncertainties and SNRs are also computed based on empirical estimates taken from neighboring frequency windows of the same width. 

\def\imin{i_{\rm min}}
\def\imax{i_{\rm max}}
\def\nwin{n_{\rm win}}
\def\Rk{R_k}
\vspace{8pt}

\noindent To determine the selection window for each SFT, the following algorithm is used. First, the template bin $i_0$ with maximum expected signal power is found, and the frequency window of minimum / maximum indices $\imin$ / $\imax$ are initialized to $\imin=\imax=i_0$. In an iterative process, $\imin$ and $\imax$ are adjusted until the fraction $\Rk$ of total window power is at least 99\%\ of the total signal power in the SFT, that is, until


\begin{equation}
    \Rk = \frac{\sum_{i=\imin}^{\imax} \Ski}{\sum_{{\rm all}\, i} \Ski} \ge 0.99.
\end{equation}

\def\hinj{h_{\rm Inj}}
\noindent At each iteration $\imin$ is decremented by one or $\imax$ incremented by 1, depending on which bin yields a greater increase in signal template power. For later convenience, we also define a Fourier transform conversion factor $\Lambda_k$:


\begin{equation}
    \Lambda_k = \frac{\sum_{i=\imin}^{\imax} \Ski}{\hinj^2}.
\end{equation}

\def\Pk{P[k]}
\def\Pkmctrl{P_{\rm ctrl}[m,k]}
\def\RPk{\Re\{P[k]\}}
\def\IPk{\Im\{P[k]\}}

\noindent where $\hinj$ is the template signal amplitude used in the time series from which $\ski$ is computed. Once the selection window is defined for SFT $k$, a (complex) product sum $\Pk$ is defined from the inner product of the
data with the template in that window:

\begin{equation}
  \Pk = \sum_{i=\imin}^{\imax} \dki (\ski)^*,
\end{equation}

\noindent where, ideally, $\RPk$ contains the injected signal and has an expectation value proportional to $h_0$, and
the expectation value of $\IPk$ is zero. If, however, the signal injection has an incorrect phase offset (from
imperfect actuation or from an imperfect calibration), then the expectation value of $\IPk$ can be non-zero.

\def\hk{h_{\rm est}[k]}
\def\dphi{\Delta\Phi}
\def\dphik{\dphi_{\rm est}[k]}
\def\hcosk{h_{\cos}[k]}
\def\hsink{h_{\sin}[k]}
\def\hcos{\hat h_{\cos}}
\def\hsin{\hat h_{\sin}}

\vspace{8pt}

\noindent To normalize these quantities to yield direct injection amplitude, we apply the scale factor $\Lambda_k$ and
correct for the template amplitude contribution to the product sum:

\begin{equation}
    \hcosk = \frac{\RPk}{\Lambda_k \hinj};
\end{equation}

\begin{equation}
    \hsink = \frac{\IPk}{\Lambda_k \hinj}.
\end{equation}

\noindent to define``cosine'' and ``sine'' projections of the data onto the template.

\vspace{8pt}

\noindent For each SFT $k$, an estimate of the injection amplitude $h_0(k)$ and phase mismatch $\dphik$ is found from

\begin{equation}
    \hk = \sqrt{(\hcosk)^2 + (\hsink)^2};
\end{equation}

\begin{equation}
    \dphik = \arctan(\hsink/\hcosk)
\end{equation}

\noindent where, ideally, $\dphik$ has an expectation value of zero.

\vspace{8pt}

\noindent As an empirical estimate of the noise contamination, the same computation is carried out
for additional, adjacent control windows of the same width as the selection window, with different bands of data multiplied against the same signal template. Specifically for a control index offset $m$, the (complex) control product sum is
\begin{equation}
  \Pkmctrl = \sum_{i=\imin}^{\imax} d[\nwin m+i] (\ski)^*,
\end{equation}

\def\varcosk{V_{\cos}[k]}
\def\varsink{V_{\sin}[k]}
\def\varcoskdef{{\rm Var}\{\Re\{\Pkmctrl\}\}}
\def\varsinkdef{{\rm Var}\{\Im\{\Pkmctrl\}\}}
\def\wcosk{w_{\cos}[k]}
\def\wsink{w_{\sin}[k]}
\def\kmax{k_{\rm max}}

\noindent where $\nwin = \imax-\imin+1$, that is, the width in bins of the selection window.
\vspace{8pt}

\noindent We determine a sample of 10 control values using $m = -5\ldots5$, excluding $m=0$, and from them
compute the variances of the cosine and sine components $\varcosk\equiv\varcoskdef$ and $\varsink\equiv\varsinkdef$ separately.
The variances are then used to compute weights:

\begin{equation}
    \wcosk = \frac{1}{\varcosk}; \qquad \wsink = \frac{1}{\varsink}.
\end{equation}

\noindent These weights are used, in turn, to compute cumulative best estimates of
cosine and sine projections
from all SFTs up to and including SFT $\kmax$:

\begin{equation}
    \hcos = \frac{\sum_{k=1}^{\kmax} \wcosk \hcosk}{\sum_{k=1}^{\kmax} \wcosk};
\end{equation}

\begin{equation}
    \hsin = \frac{\sum_{k=1}^{\kmax} \wsink \hsink}{\sum_{k=1}^{\kmax} \wsink}.
\end{equation}

\def\sigmahcos{\sigma^{\hat h}_{\cos}}
\def\sigmahsin{\sigma^{\hat h}_{\sin}}
\noindent which have estimated uncertainties:

\begin{equation}
    \sigmahcos = \sqrt{\frac{1}{\sum_{k=1}^{\kmax} \wcosk}};
\end{equation}

\begin{equation}
    \sigmahsin = \sqrt{\frac{1}{\sum_{k=1}^{\kmax} \wsink}};
\end{equation}

\noindent As for individual SFTs, one can compute cumulative injection amplitude and phase mismatch from these quantities:


\begin{equation}
    h_0 = \sqrt{(\hcos)^2 + (\hsin)^2};
\end{equation}

\begin{equation}
    \dphi = \arctan(\hsin/\hcos).
\end{equation}

\noindent which have estimated uncertainties:

\begin{equation}
    \sigma_{h_0} =
    \sqrt{\frac{(\hcos)^2(\sigmahcos)^2 + (\hsin)^2(\sigmahsin)^2}
    {(\hcos)^2 + (\hsin)^2}};
\end{equation}

\begin{equation}
    \sigma_{\dphi} =
    \frac{\sqrt{(\hcos)^2(\sigmahcos)^2 + (\hsin)^2(\sigmahsin)^2}}
    {(\hcos)^2 + (\hsin)^2};
\end{equation}

\noindent The SNR is computed from:

\begin{equation}
    {\rm SNR} = \frac{h_0}{\sigma_{h_0}}.
\end{equation}

\noindent Example graphs of recovered injection properties \vs\ time are presented in section~\ref{sec:template_results}, along with summary
graphs for all injections.

\subsection{\texorpdfstring{\Fstatistic}{F-statistic} Method}

The \Fstatistic-based monitoring exploits well established LIGO Analysis Library ({\tt lalsuite})~\cite{LALSuite_2018,} software to
compute both daily and cumulative \Fstatistic~\cite{Jaranowski_1998} values for each day of the run for each interferometer.
In addition to computing an \Fstatistic\ value (denoted by \fstatistic) for each injection,
interferometer and time span, the program 
(lalpulsar\_ComputeFstatistic\_v2~\cite{Prix_2009_fstat}) computes best estimates and uncertainties for
the injection amplitude $h_0$, cosine of the inclination angle $\iota$, phase offset $\phi_0$ and polarization angle $\psi$.

\vspace{8pt}

\noindent In brief, the \Fstatistic\ is constructed from a combination of functions computed numerically from the
SFTs of the observation span. These functions take into account the antenna pattern responses of the
detector at the midpoints of the SFTs, and their combination is maximized with respect to $h_0$, $\cos(\iota)$, $\phi_0$ and $\psi$.
In the \fstatistic\ implementation, the
source spindown is fixed at the intended injected value, but the frequency is allowed to vary over a
narrow band around the intended value. The original \Fstatistic\ definition may be found in~\cite{Jaranowski_1998}, and
the algorithm implementation in the {\tt lalsuite} package may be found in~\cite{Prix_2009_fstat}.

\vspace{8pt}

\noindent Analytically maximizing \fstatistic\ with respect to the four intrinsic source parameters allows solving for the
best estimates for those parameters, with the Fisher matrix used in the maximization process implying estimated
uncertainties. These quantitities are used in daily produced graphs.

\vspace{8pt}

\noindent Specifically, for each injection monitored and for each interferometer, graphs are produced for
1) the \fstatistic\ for that day's data; 2) the cumulative \fstatistic\ from the start of the run through that day; and 3-6) cumulative graphs of the four estimated signal
parameters \vs\ time. Examples are shown in Section~\ref{sec:fstatresults}.

\subsection{Bayesian Method}\label{sec:bayesian_method}

The Bayesian method \cite{Dupuis_2005, Pitkin_2017} starts by greatly compressing the calibrated gravitational-wave
strain data and then uses this compressed data, and the assumption of stationary noise with varying timescales, to
estimate posterior probabilities on the unknown signal parameters: $h_0$, $\iota$, $\psi$, and $\Phi_0$. The method
is able to coherently analyze any length of observed data provided the phase evolution can be accurately tracked over
the observation period, which is the case for many known pulsars and certainly the case for the hardware injections.
For the compression, for each hardware injection the phase evolution from Eqn.~\ref{eq:phase_evolution} is calculated
over the observation period. The strain data is multiplied by $e^{-i\Phi(t)}$ (with the value of $\Phi_0$ not
included), i.e., heterodyned, which produces a complex time series
in which the signal power is equally split between a component now shifted to $\sim 0$\,Hz (D.C.; with small modulations
left in from the diurnal antenna patterns), and one shifted to twice the signal frequency. The noise at the signal
frequency is also equally split in the same way, meaning the signal-to-noise ratio for each component is the same
as the original signal. This complex time series is then heavily low-pass filtered using a 9$^{\rm th}$ order
Butterworth IIR filter with a knee-frequency of 0.25\,Hz, and down-sampled (via averaging) from a sample rate of
16\,384\,Hz to 1/60\,Hz. The resulting heterodyned complex time series contains approximately white Gaussian noise over a
bandwidth of
$\pm 1/120$\,Hz about the signal, which now has the form

\begin{eqnarray}
    h'(t) = \frac{h_0}{2}\left(\frac{1}{2}F_{+}(t, \psi)(1+\cos{}^2\iota)e^{i\Phi_{0}}
    -iF_{\times}(t, \psi)\cos{\iota}e^{i\Phi_{0}} \right).
\end{eqnarray}

\noindent where $\Phi_{0}$ is defined in Eqn.~\ref{eq:phase_evolution}.
To infer the signal parameters for each injection, their joint posterior probability distribution is calculated
via Bayes' rule

\begin{equation}
    p(h_0, \cos{\iota}, \Phi_0, \psi|\mathbf{D}) \propto p(\mathbf{D}| h_0, \cos{\iota}, \Phi_0, \psi) p(h_0, \cos{\iota}, \Phi_0, \psi),
\end{equation}

\vspace{8pt}

\noindent where $\mathbf{D}$ is the heterodyned time series from a given detector and for a particular injection, $p(\mathbf{D}| h_0, \cos{\iota}, \Phi_0, \psi)$ is the likelihood function of the data, and $p(h_0, \cos{\iota}, \Phi_0, \psi)$ is the joint prior probability
for the parameters. In this analysis, the priors are treated as a constant, i.e., uniform distributions over the allowed
parameter ranges. The likelihood function assumes Gaussian noise with an unknown variance over a series of stationary chunks. The length of each stationary chunk is calculated via change-point analysis using a method based on {\it BayesianBlocks} \cite{Scargle_1997, Pitkin_2017}. The unknown variance in each chunk can be marginalized over, to produce a Student's t-likelihood \cite{Dupuis_2005}. The full likelihood is then calculated as the product of the likelihoods for each chunk.

\vspace{8pt}

\noindent In practice, the posterior probability distribution is sampled from using the Nested Sampling method \cite{Ashton_2022, Pitkin_2017, Pitkin_2022}, giving a series of samples $\{h_0, \cos{\iota}, \psi, \Phi_0\}$ that are distributed according to the posterior and from which credible intervals from the marginal distributions for each parameter can be calculated.

\section{Results}

Here we present example and summary results from the three complementary monitoring programs. For simplicity, most results are shown for strain data not subjected to self-gating of loud transients, but similar and consistent results are observed for corresponding self-gated data (see below).

\subsection{Template Generation Method}
\label{sec:template_results}
\subsubsection{Differential Analysis of Weak and Strong Injections} \hfill

\noindent The analysis focuses on recovering and comparing key parameters, including amplitude, phase, and frequency evolution, for both weak and strong injections. This approach involves assessing the consistency of recovered strain amplitudes across signals of different strengths, evaluating the precision of phase recovery to identify biases between weak and strong signals, and examining variations in the SNR as a function of signal strength to gain insights into the detector's performance and reliability~\cite{Covas_PRD_2018}
.

\vspace{8pt}

\noindent Diagnostic plots shown below are used to visualize trends and discrepancies in these metrics. For weak signals, the proximity to the noise floor may result in greater variability in recovered parameters, whereas strong signals exhibit higher recovery accuracy. These visualizations help identify performance boundaries and calibration inconsistencies.

\vspace{8pt}

\begin{figure}[ht]
\centering
\begin{subfigure}[b]{0.48\linewidth}
    \centering
    \includegraphics[width=\linewidth]{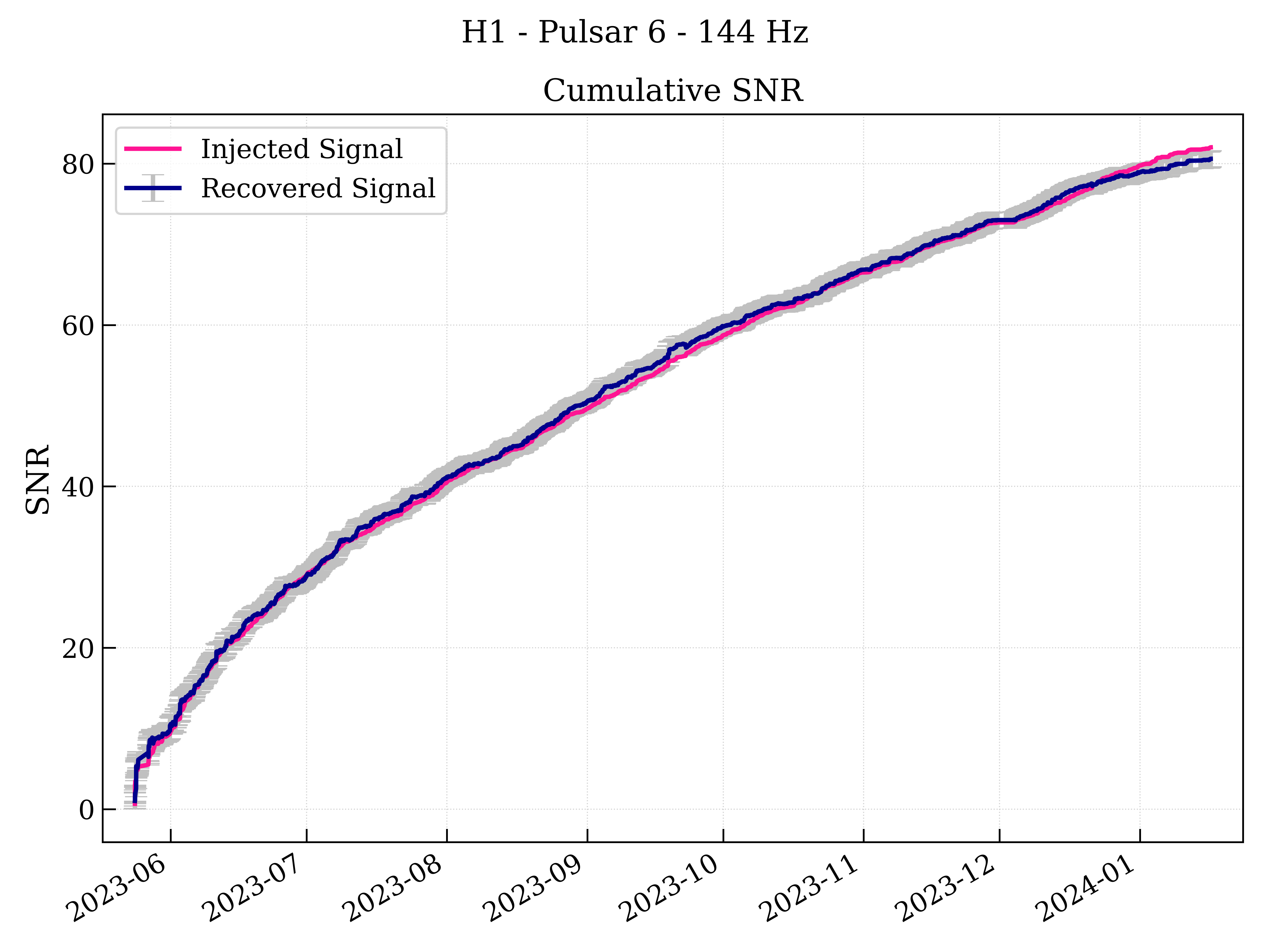}
    \label{fig:H1_SNR_6}
\end{subfigure}
\hfill
\begin{subfigure}[b]{0.48\linewidth}
    \centering
    \includegraphics[width=\linewidth]{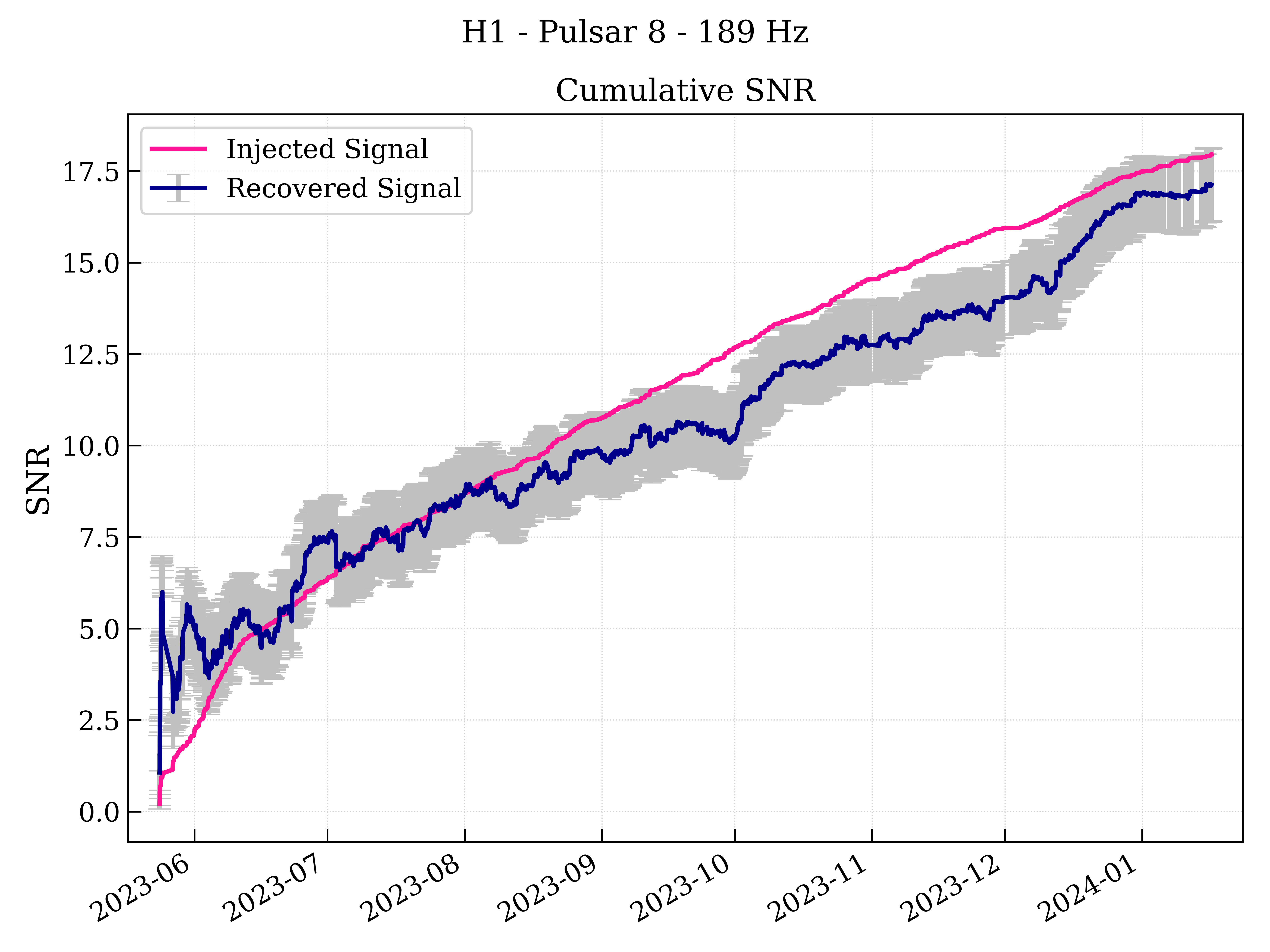}
    \label{fig:H1_SNR_8}
\end{subfigure}

\begin{subfigure}[b]{0.48\linewidth}
    \centering
    \includegraphics[width=\linewidth]{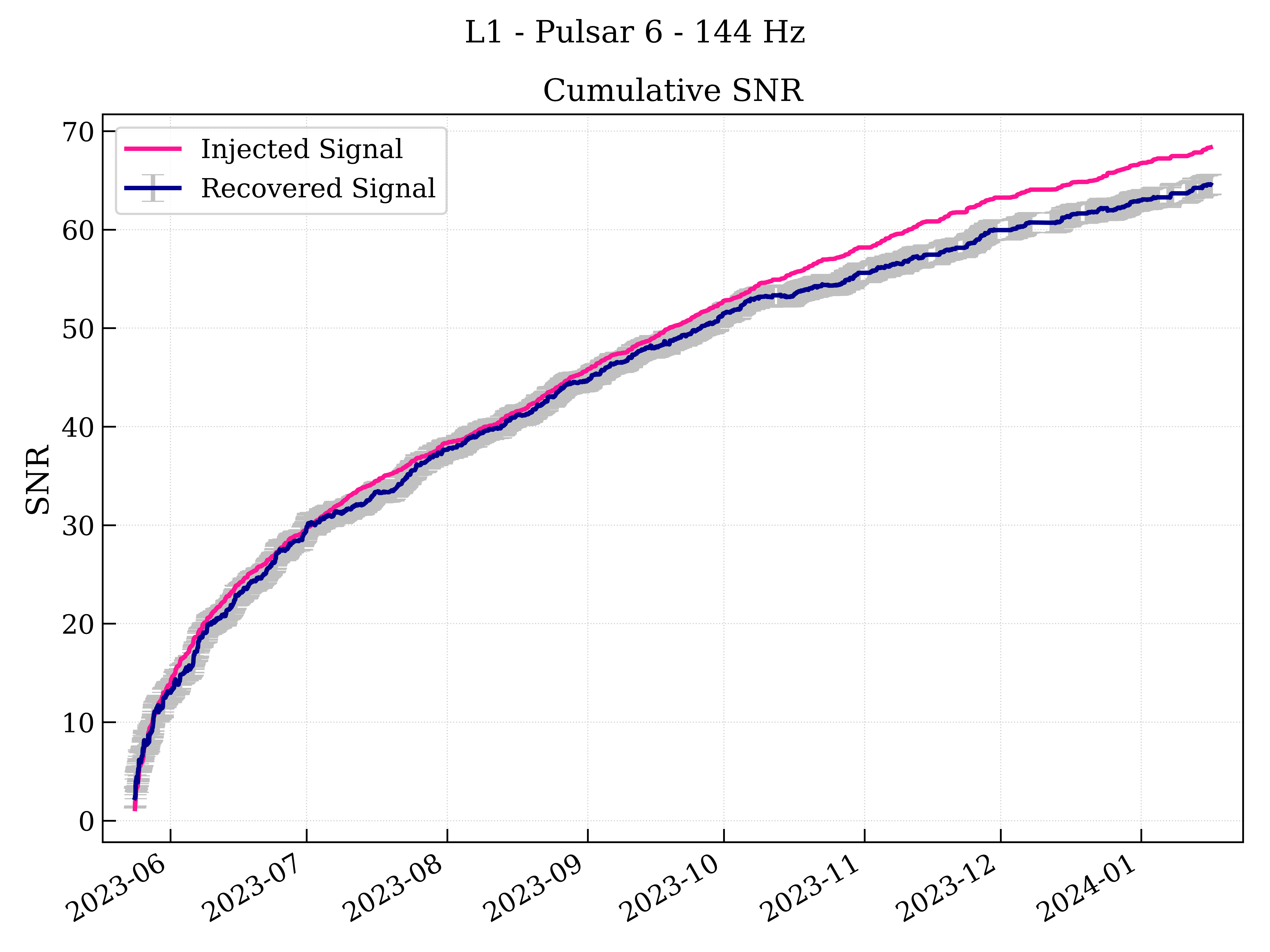}
    \label{fig:L1_SNR_6}
\end{subfigure}
\hfill
\begin{subfigure}[b]{0.48\linewidth}
    \centering
    \includegraphics[width=\linewidth]{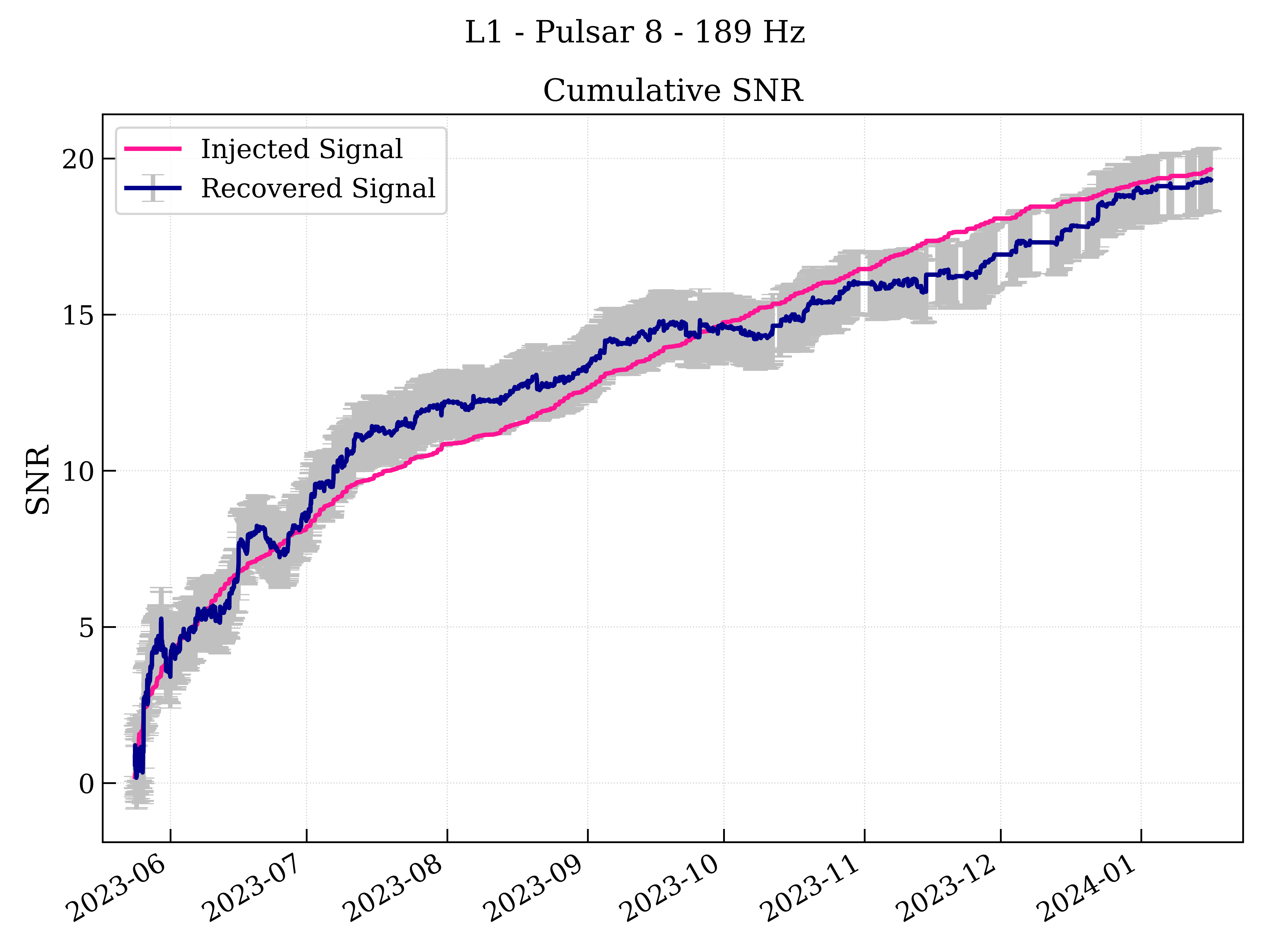}
    \label{fig:L1_SNR_8}
\end{subfigure}

\caption{Cumulative SNR measurements for Injection 6 and Injection 8 from H1 and L1 detectors. The first row corresponds to H1 data, while the second row represents L1 data. Adjacent error bars in these cumulative
graphs are highly correlated.}
\label{fig:combined_summary_SNR}
\end{figure}

\noindent Sample results from the cumulative SNR monitoring of Injection 6 ($\sim$144 Hz) and Injection 8 ($\sim$189 Hz) injections in both H1 and L1 are shown in Figure~\ref{fig:combined_summary_SNR}. In each panel the deep pink curve is the expected $\sqrt{T}$ growth of the injected SNR for the intended injection amplitude, the dark blue curve is the daily recovered SNR from our narrow-band search, and the gray bars show the $\pm1\sigma$ uncertainty across the frequency band.

\vspace{8pt}

\noindent One sees in H1 for Injection 6 (upper left panel) excellent agreement between expected and actual
rise in SNR, but L1 (lower left) there is some discrepancy by the end of the O4a period. In
these cumulative graphs, adjacent SNR values and their error bars shown are highly correlated
since nearly the same data is used to evaluate them.
For the weaker Injection 8
(upper and lower right), both detectors exhibit larger early fluctuations as the estimator initially latches onto noise excursions, but by the end of O4a, there is good agreement between expectation and signal recovery.

\begin{figure}[ht]
\centering
\begin{subfigure}[b]{0.49\linewidth}
    \centering
    \includegraphics[width=\linewidth]{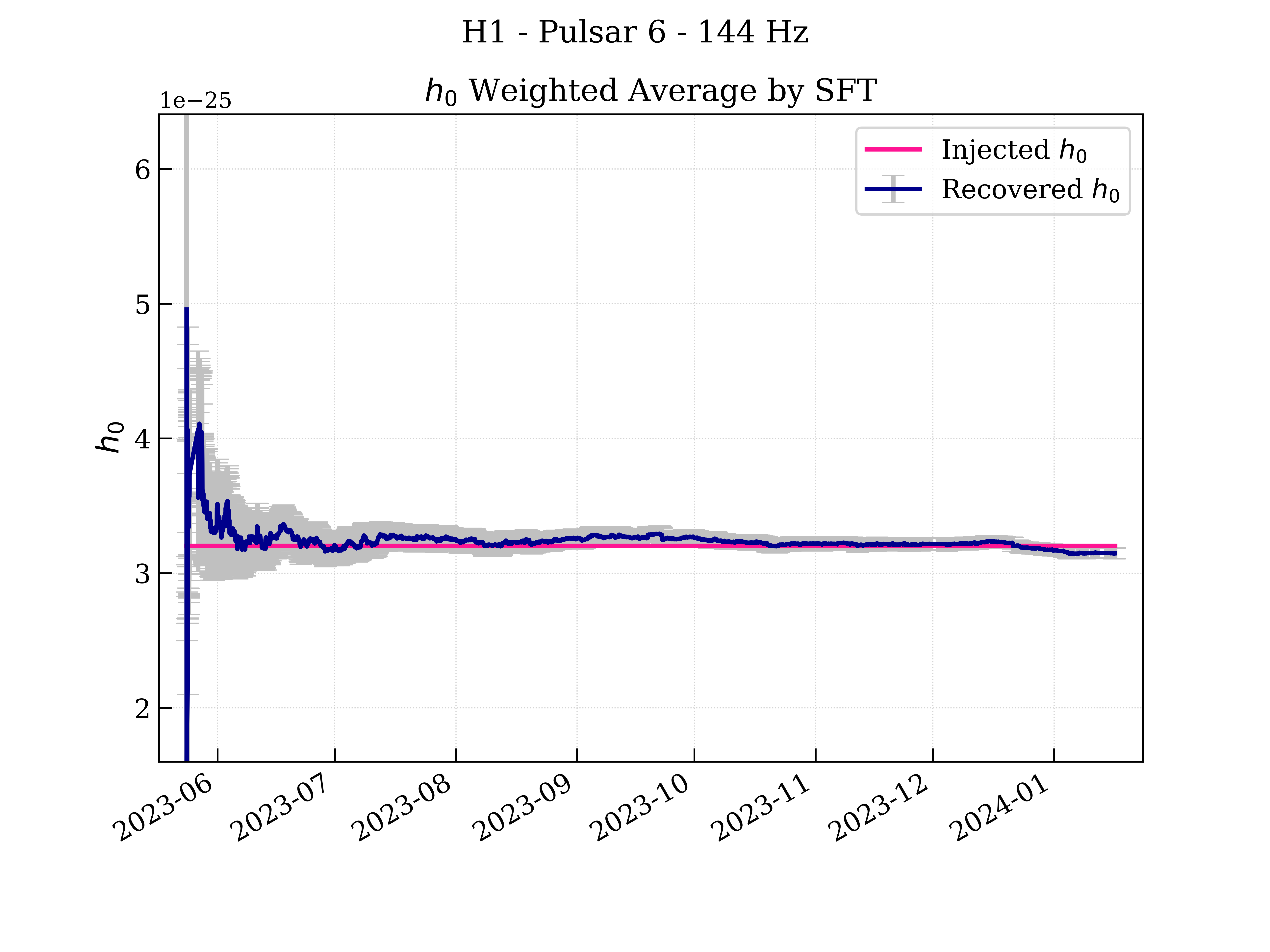}
    \label{fig:H1_weighted_avg_1}
\end{subfigure}
\hfill
\begin{subfigure}[b]{0.49\linewidth}
    \centering
    \includegraphics[width=\linewidth]{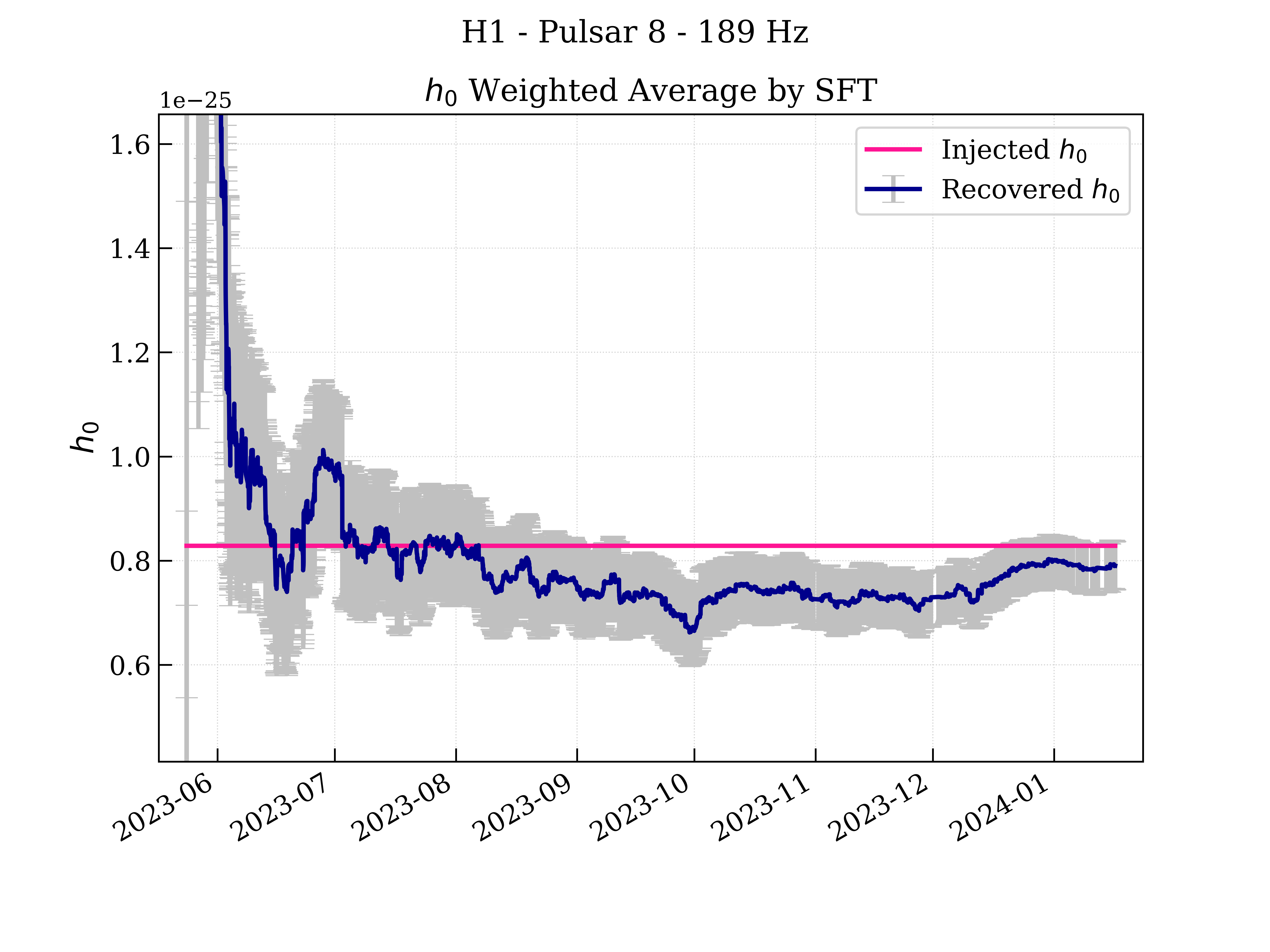}
    \label{fig:H1_weighted_avg_2}
\end{subfigure}

\begin{subfigure}[b]{0.49\linewidth}
    \centering
    \includegraphics[width=\linewidth]{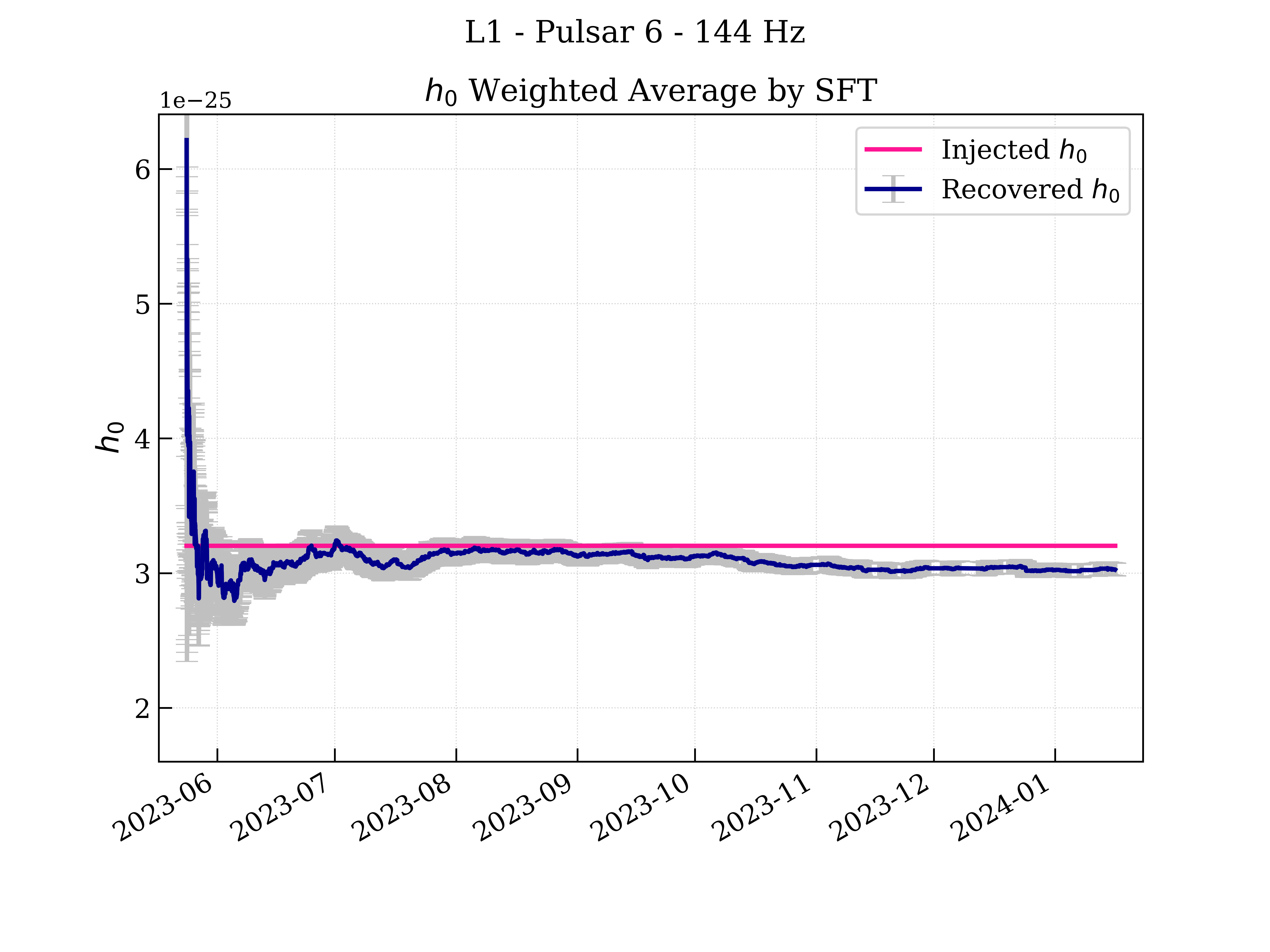}
    \label{fig:L1_weighted_avg_1}
\end{subfigure}
\hfill
\begin{subfigure}[b]{0.49\linewidth}
    \centering
    \includegraphics[width=\linewidth]{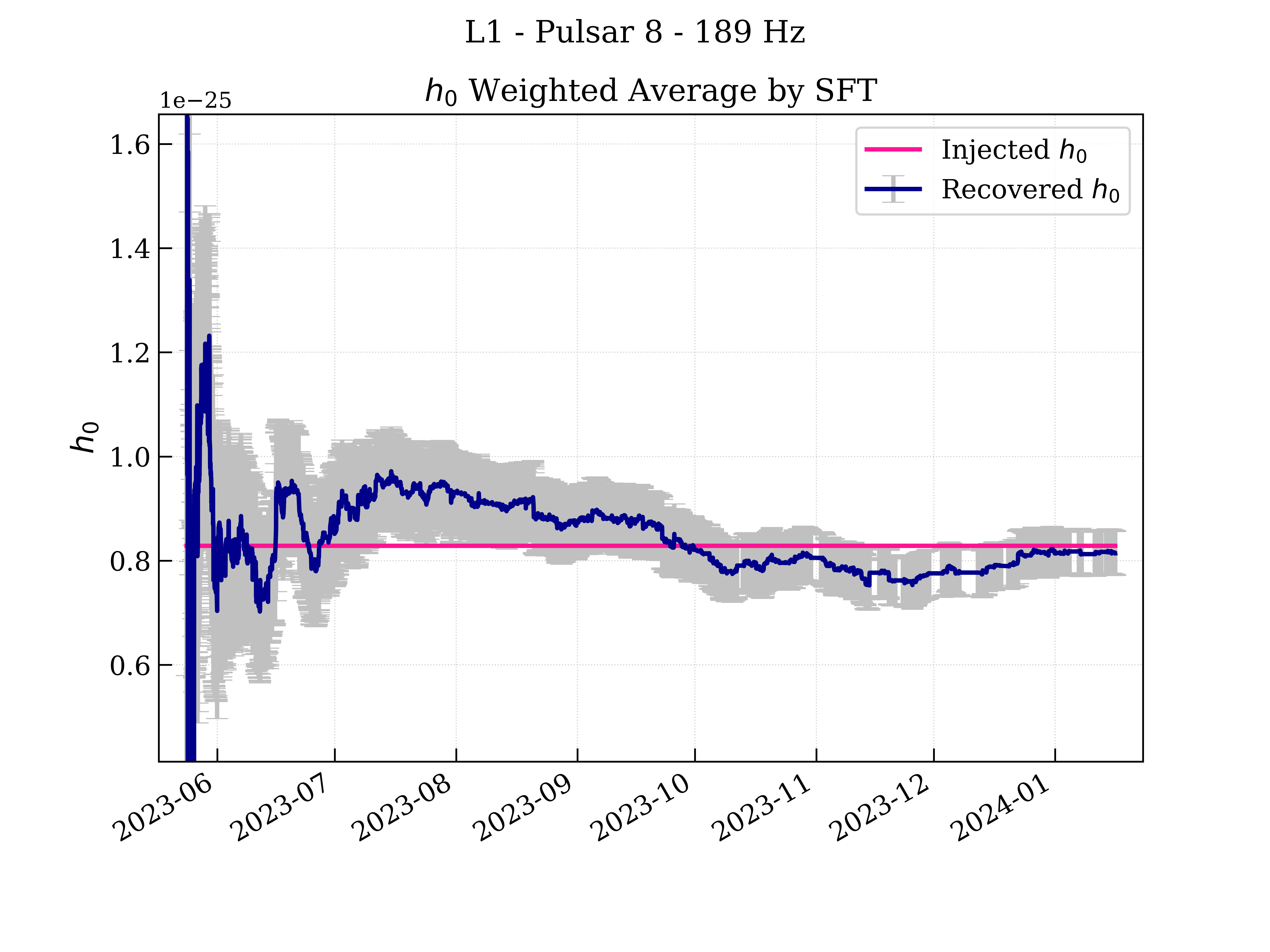}
    \label{fig:L1_weighted_avg_2}
\end{subfigure}

\caption{Weighted average \( h_0 \) by SFT for Injection 6 and Injection 8. The first row shows H1 data, while the second row represents L1 data.}
\label{fig:weighted_avg_combined}
\end{figure}

\noindent Figure~\ref{fig:weighted_avg_combined} presents the SFT‐weighted estimates of the gravitational‐wave amplitude \(h_0\) for Injection 6 ($\sim$144 Hz) and Injection 8 ($\sim$189 Hz) in the H1 (top row) and L1 (bottom row) detectors. In each plot the deep pink curve marks the true injected \(h_0\), the dark blue trace the recovered weighted average, and the gray error bars the \(\pm1\sigma\) spread across individual SFT estimates. For the strong Injection 6, both H1 and L1 show an initial noise‐driven overshoot in early June, but by mid‐June the recovered \(h_0\) converges to within a few percent of the intended injected value ( in both H1 and L1), and the uncertainties shrink as more SFTs accumulate. A small discrepancy is seen in the reconstructed L1 amplitude for Injection 6, but consistent with a statistical fluctuation from the collective
mean reconstructed / intended amplitude (see Fig.~\ref{fig:relative_amplitudes_combined}).
By contrast, the weaker Injection 8 exhibits large early scatter—error bars exceeding 30\% of the injected amplitude — and a pronounced bias as the maximization briefly latches onto statistical fluctuations. Over the next three months, however, both detectors’ weighted averages drift steadily toward the true \(h_0\) and their $1\sigma$ bands contract, achieving agreement within the error envelope by September 2023 in H1 and by October in L1. The fluctuations early in the run for weak injections merely


\begin{figure}[ht]
\centering
\begin{subfigure}[b]{0.48\linewidth}
    \centering
    \includegraphics[width=\linewidth]{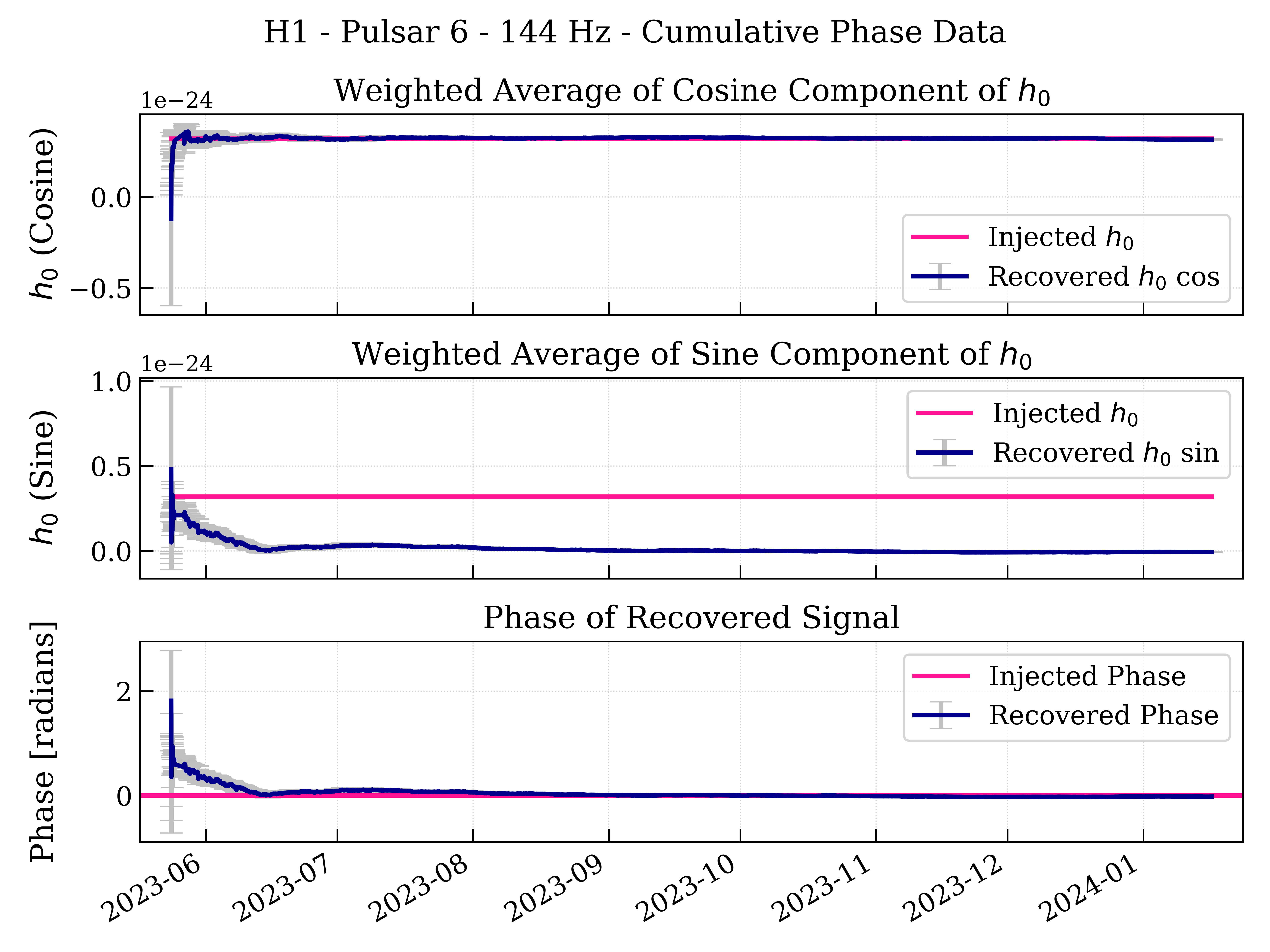}
    \label{fig:H1_cumulative_phase_1}
\end{subfigure}
\hfill
\begin{subfigure}[b]{0.48\linewidth}
    \centering
    \includegraphics[width=\linewidth]{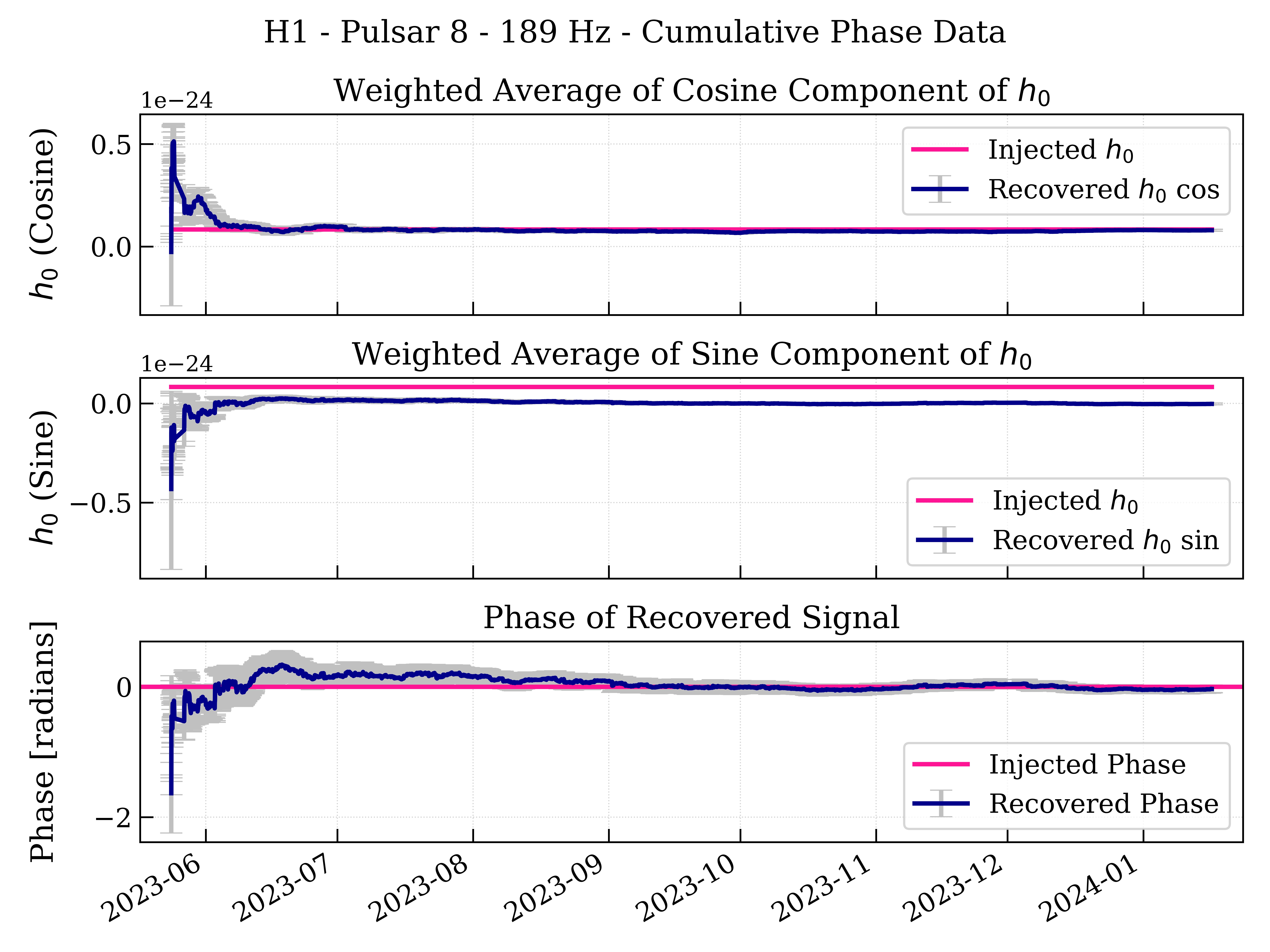}
    \label{fig:H1_cumulative_phase_2}
\end{subfigure}

\begin{subfigure}[b]{0.48\linewidth}
    \centering
    \includegraphics[width=\linewidth]{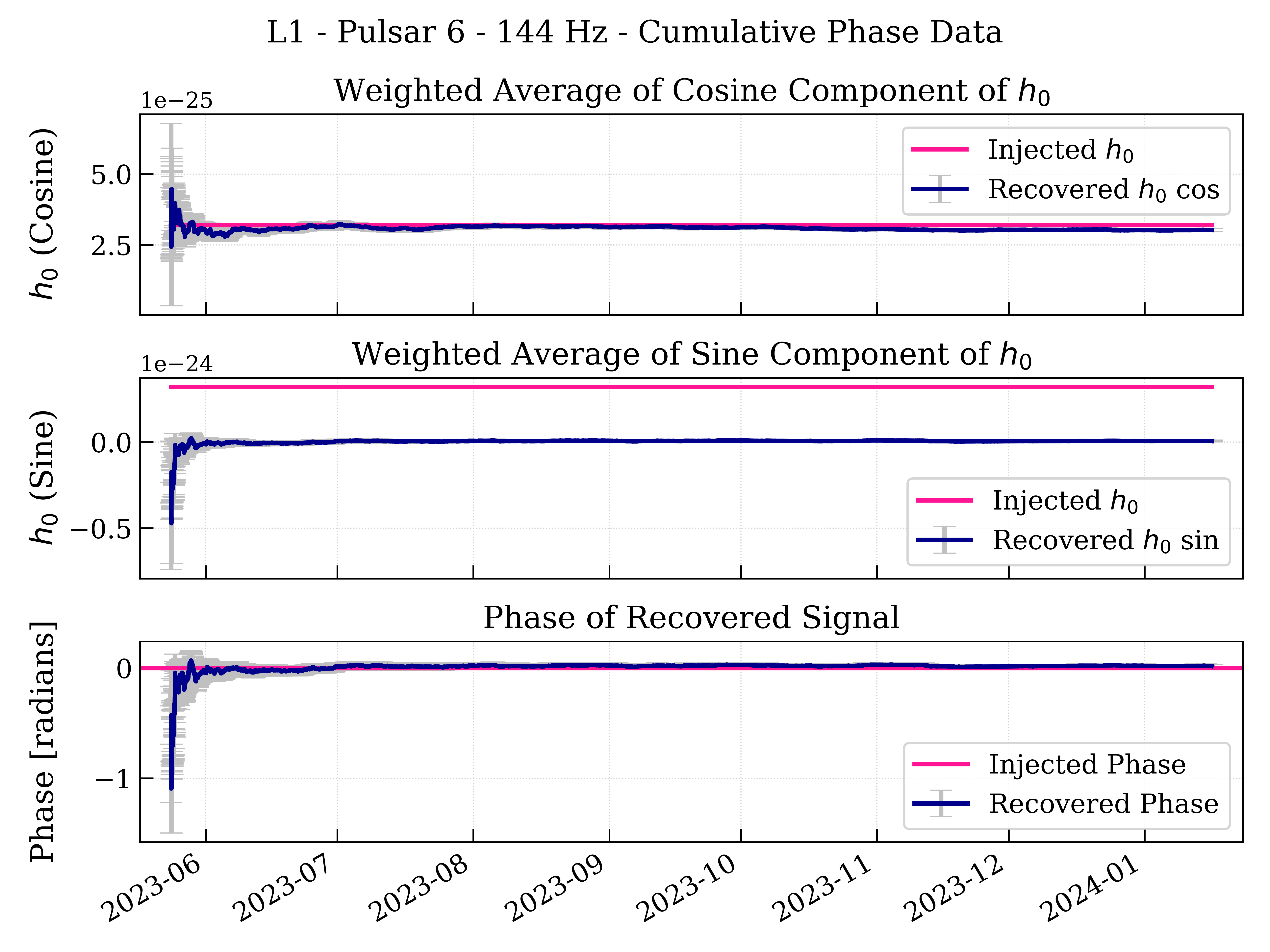}
    \label{fig:L1_cumulative_phase_1}
\end{subfigure}
\hfill
\begin{subfigure}[b]{0.48\linewidth}
    \centering
    \includegraphics[width=\linewidth]{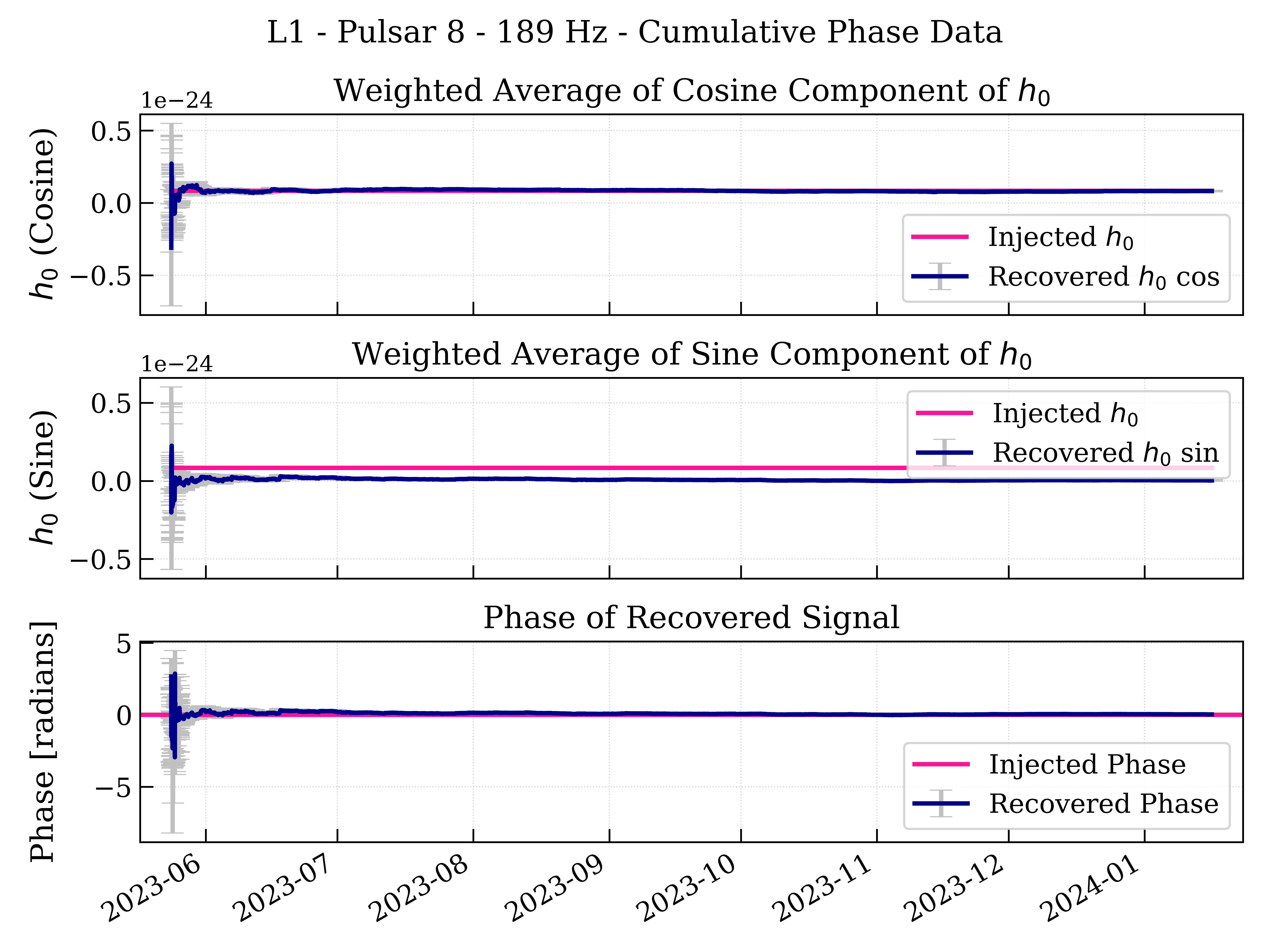}
    \label{fig:L1_cumulative_phase_2}
\end{subfigure}

\caption{Cumulative phase offset reconstructions for Injection 6 (left panels) and Injection 8 (right panels) from the H1 and L1 detectors. The top panels correspond to H1 data, while the bottom panels depict L1 data. Within each panel, the top graph shows the cosine projection of the reconstructed amplitude, the middle graph shows the sine projection, and the bottom graph shows the inferred phase offset. Ideally, the cosine projection should agree with the intended injected $h_0$ shown, and the sine projection should be consistent with zero. The intended $h_0$ is shown on the sine graphs for scale.}
\label{fig:cumulative_phase_combined}
\end{figure}

\noindent Figure~\ref{fig:cumulative_phase_combined} illustrates the cumulative phase offset reconstruction for Injection 6 ($\sim$144 Hz) and Injection 8 ($\sim$189 Hz) in the H1 (top row) and L1 (bottom row) detectors.  In each column the upper panel shows the SFT‐weighted average of the cosine component of \(h_0\), the middle panel the sine component, and the lower panel the recovered signal phase, all compared against the injected references (deep pink curves) with their \(\pm1\sigma\) spreads (gray bars).  For the strong Injection 6, both H1 and L1 exhibit an early noise‐driven overshoot in the cosine term that settles to the true \(h_0\) within a few weeks, while the sine component decays to near zero and the recovered phase (dark blue) rapidly locks onto the injected phase with shrinking uncertainties.

\vspace{8pt}

\noindent In contrast, the weaker Injection 8 shows larger initial scatter in both cosine and sine components -- particularly in L1 -- accompanied by broader phase error bars; over the course of O4a the weighted averages converge toward the intended injected values, and the recovered phase offset agrees with the intended value.


\begin{figure}[ht]
\centering
\begin{subfigure}[b]{0.48\linewidth}
    \centering
    \includegraphics[width=\linewidth]{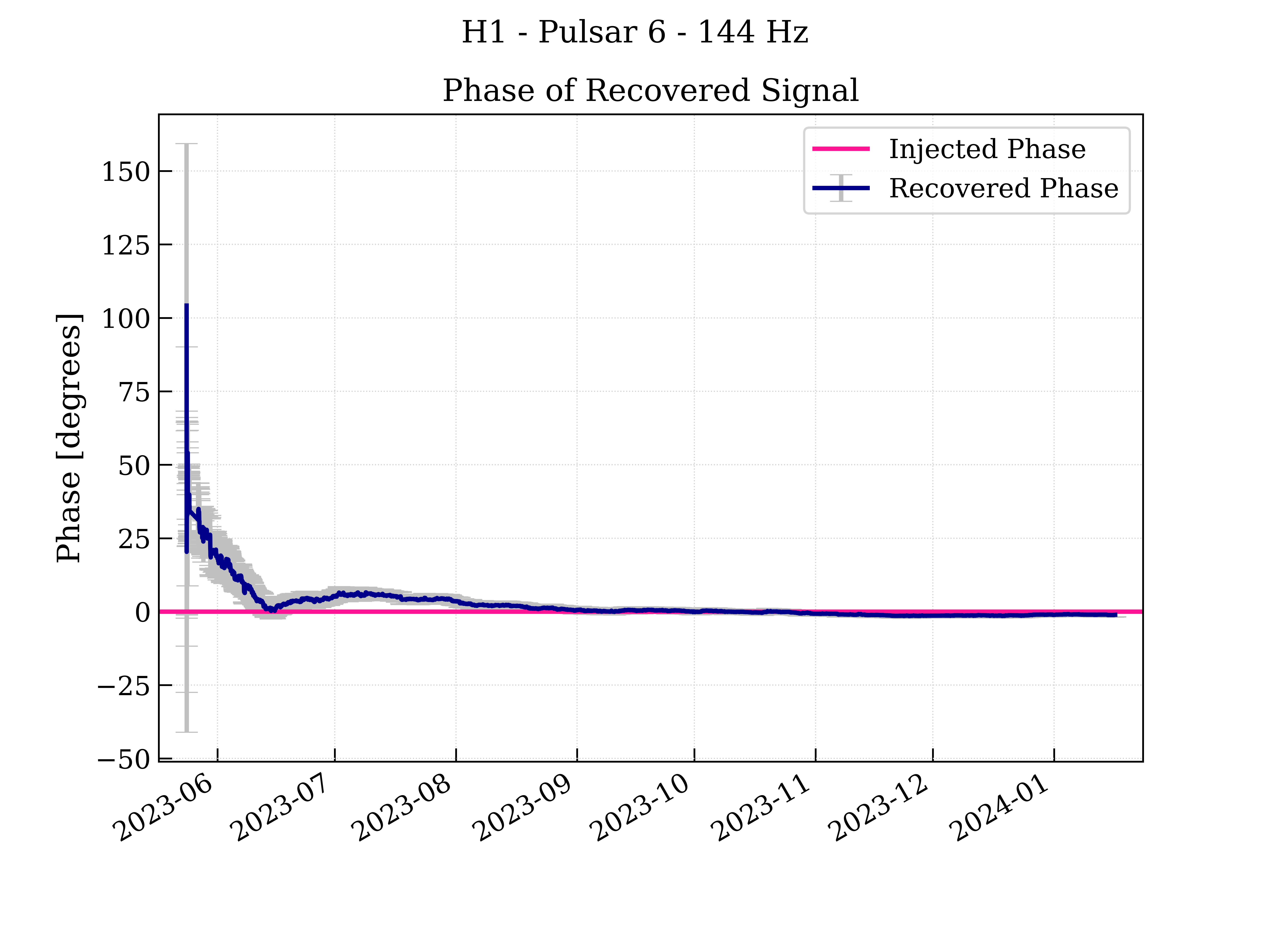}
    \label{fig:H1_recovered_phase_1}
\end{subfigure}
\hfill
\begin{subfigure}[b]{0.48\linewidth}
    \centering
    \includegraphics[width=\linewidth]{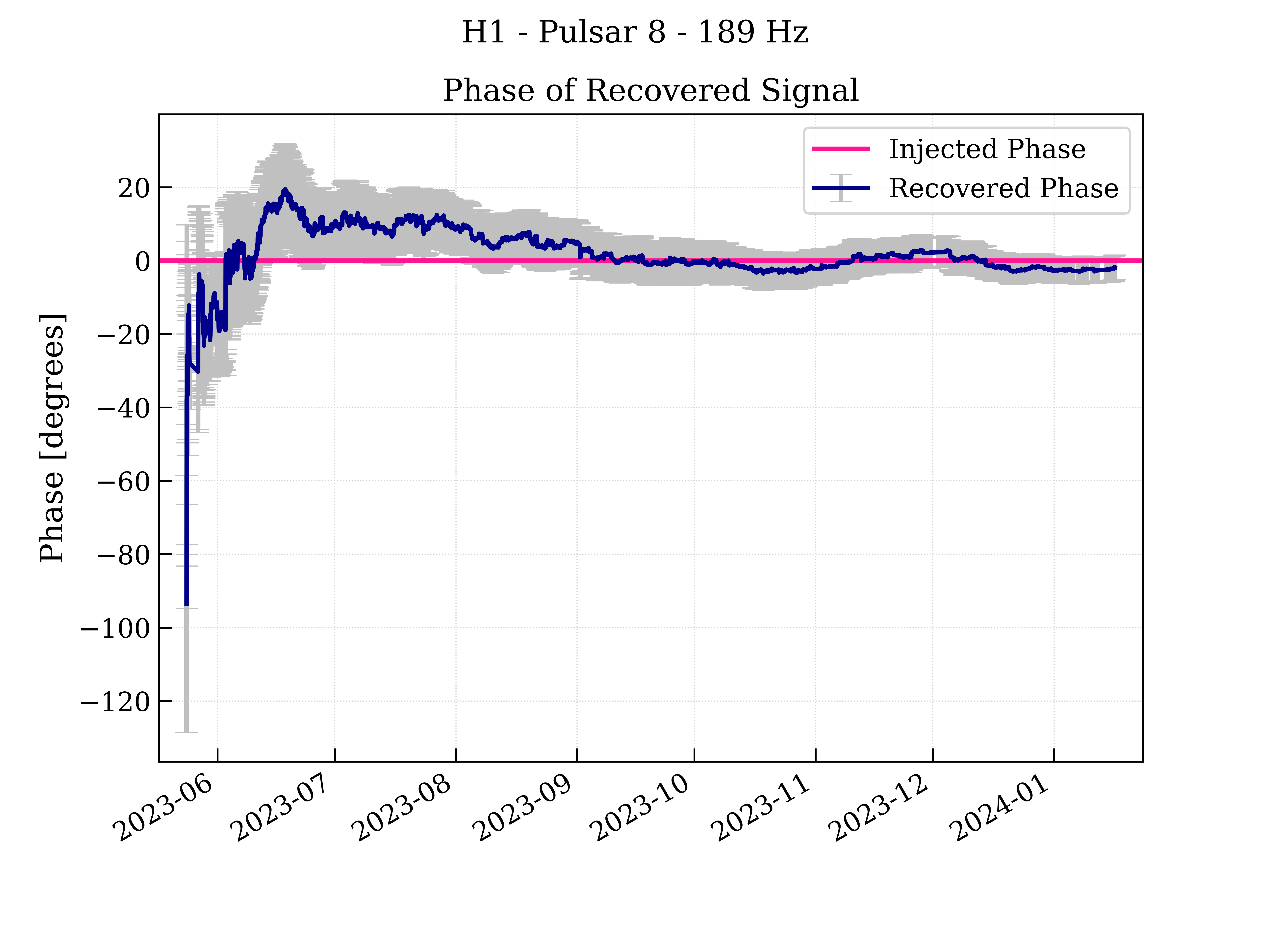}
    \label{fig:H1_recovered_phase_2}
\end{subfigure}

\begin{subfigure}[b]{0.47\linewidth}
    \centering
    \includegraphics[width=\linewidth]{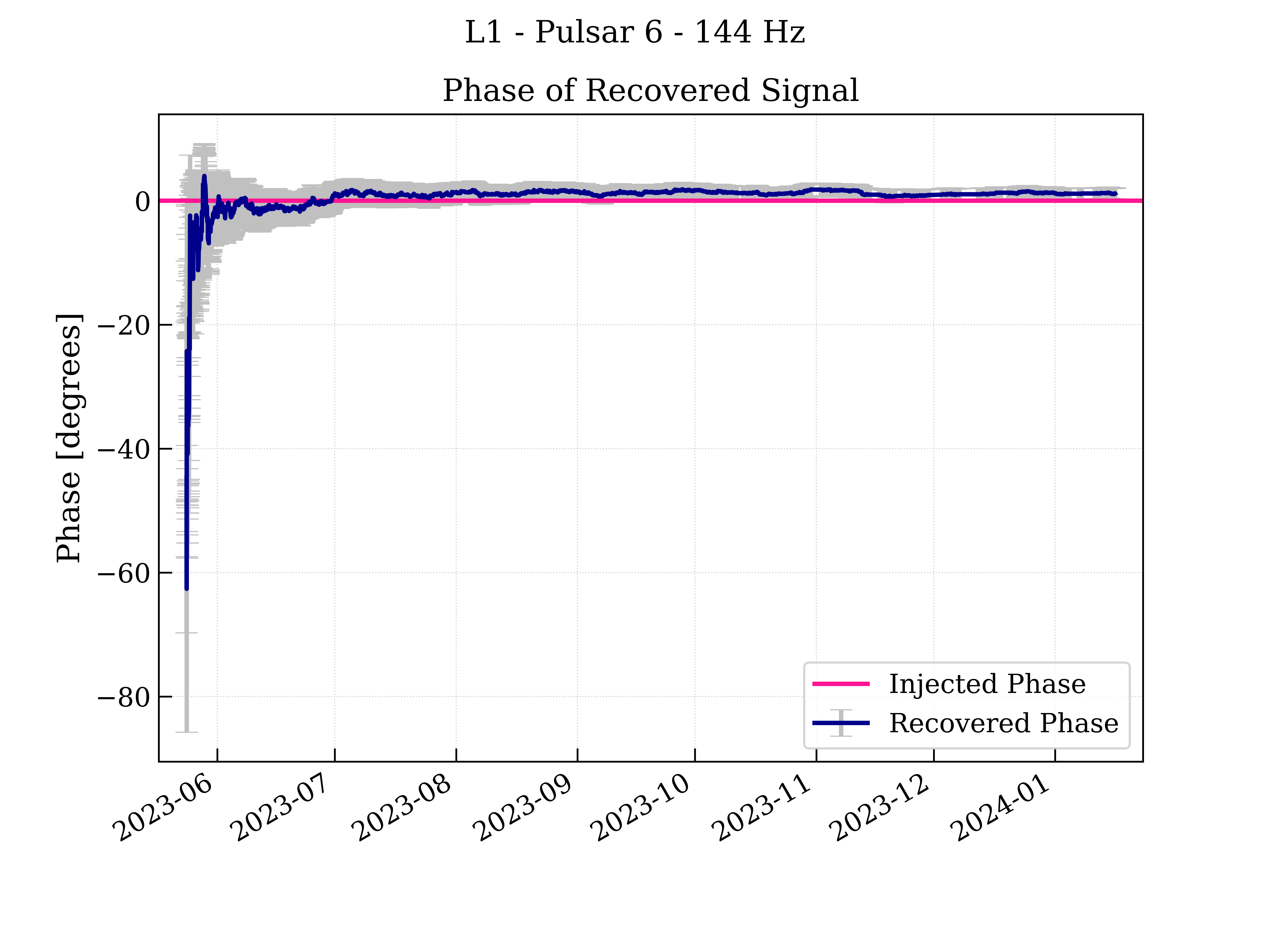}
    \label{fig:L1_recovered_phase_1}
\end{subfigure}
\hfill
\begin{subfigure}[b]{0.47\linewidth}
    \centering
    \includegraphics[width=\linewidth]{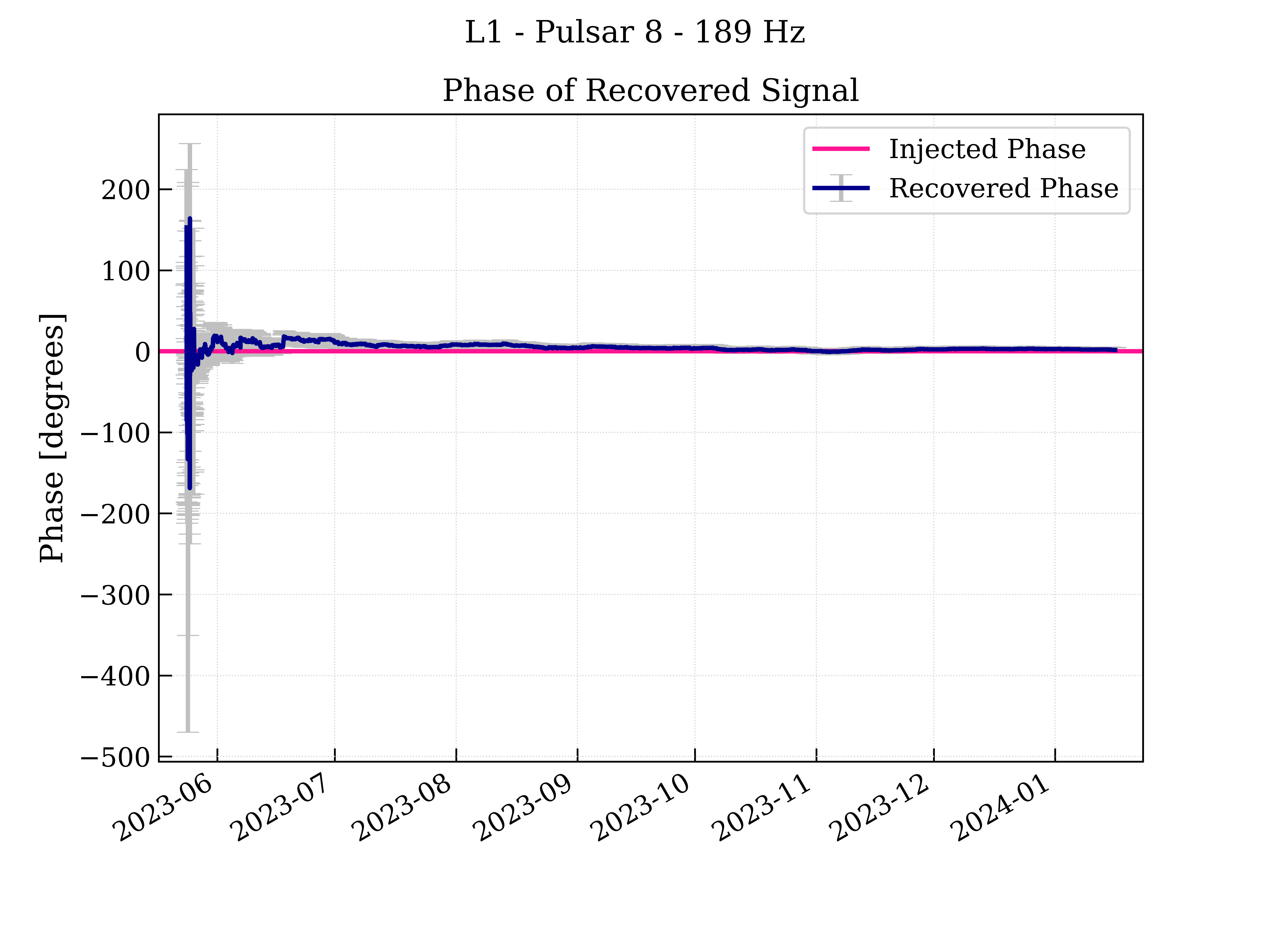}
    \label{fig:L1_recovered_phase_2}
\end{subfigure}

\caption{Relative phase offset of recovered signal for Injection 6 and Injection 8 from H1 and L1 detectors. The first row corresponds to H1 data, while the second row represents L1 data. Each column highlights the comparative performance across different metrics for the two pulsars.}
\label{fig:recovered_phase_combined}
\end{figure}

\noindent Figure~\ref{fig:recovered_phase_combined} presents the (magnified) evolution of the recovered signal relative phase offset (magenta trace with gray $\pm$1-$\sigma$ error bars) compared to the intended value (zero,
by definition -- red curve) for Injection 6 ($\sim$144 Hz) and Injection 8 ($\sim$189 Hz) in H1 (top row) and L1 (bottom row). In H1 for Injection 6 (upper left), the relative phase offset estimate exhibits a large initial scatter—spanning $\pm$150$^\circ$—as the maximization latches onto noise fluctuations, but by mid-June it rapidly converges to within a few degrees of the injected phase and remains stable thereafter. L1 for Injection 6 (lower left) shows a similar pattern with a larger early bias and broad uncertainties, converging toward
the intended value by late June.

\vspace{8pt}

\noindent For the weaker Injection 8 in H1 (upper right), the recovered relative phase offset initially undershoots (to –100°) then overshoots (+20$^\circ$) before drifting toward zero over the next three months; in L1 (lower right), the phase begins with extreme fluctuations ($\pm$200$^\circ$) and a pronounced negative bias, but gradually aligns to the injected phase by July and holds within $\pm$5$^\circ$ for the remainder of O4a.

\subsubsection{Integrated Summary of Detector Response to Injections} \hfill

\vspace{8pt}

\noindent The daily summary graphs provide a more global snapshot of H1 and L1
calibration fidelity over the full search band for CW searches, giving insight
into potential broadband discrepancies not easily seen for an individual injection.

\vspace{8pt}

\vspace{8pt}

\begin{figure}[ht]
\centering
\begin{subfigure}[b]{0.47\linewidth}
    \centering
    \includegraphics[width=\linewidth]{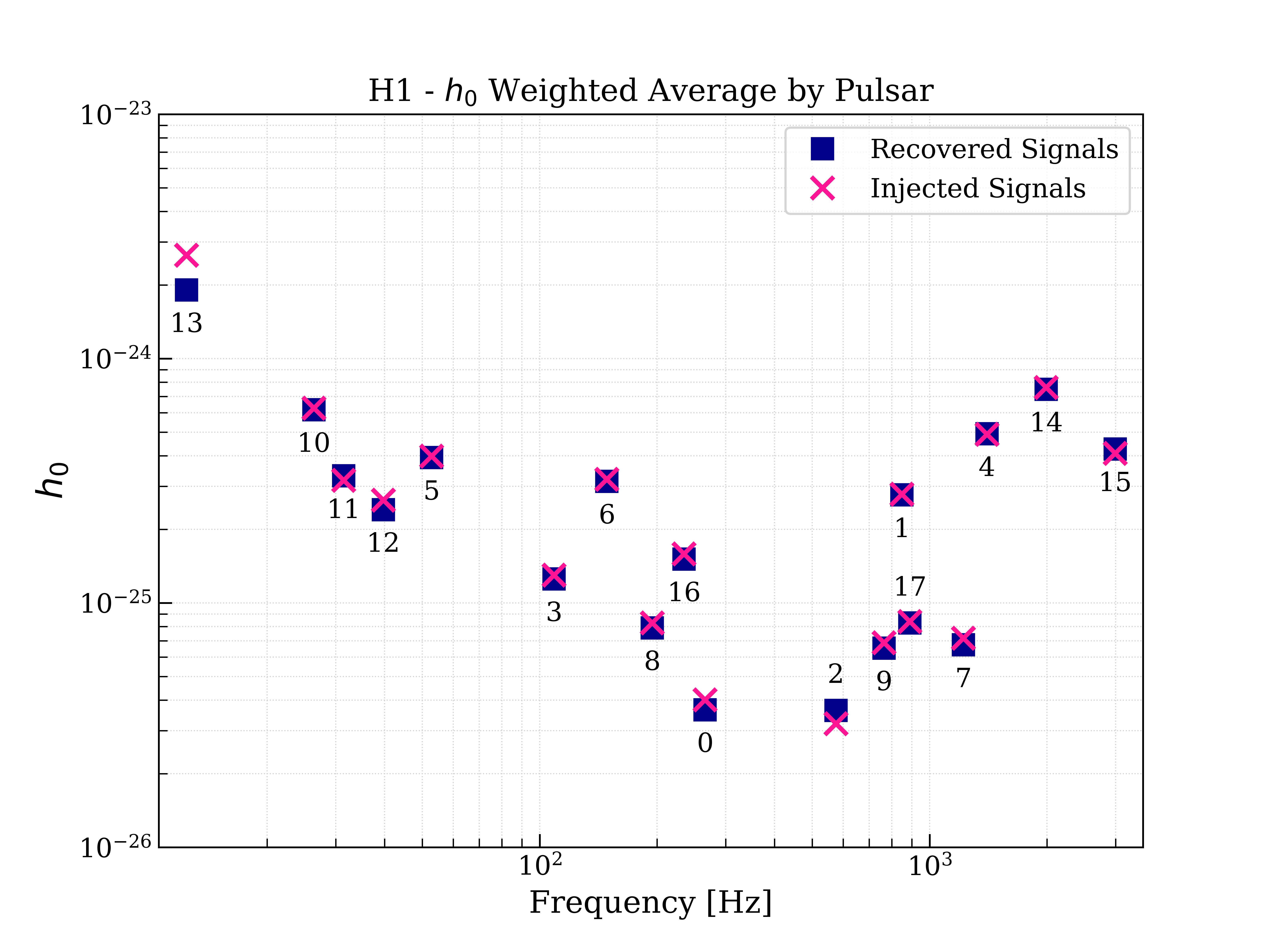}
    \label{fig:H1_weighted_avg_pulsar}
\end{subfigure}
\hfill
\begin{subfigure}[b]{0.47\linewidth}
    \centering
    \includegraphics[width=\linewidth]{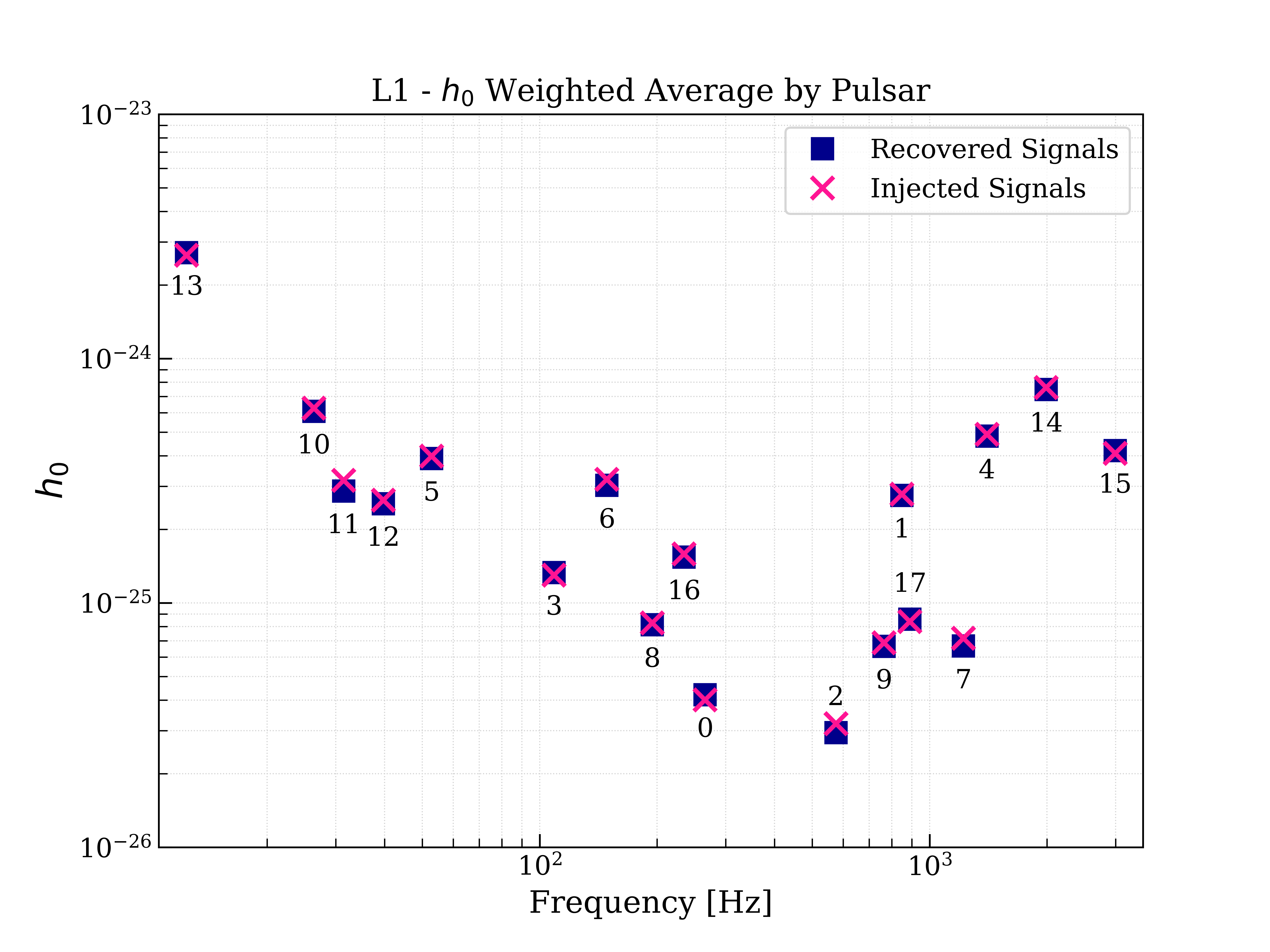}
    \label{fig:L1_weighted_avg_pulsar}
\end{subfigure}

\caption{Weighted average \( h_0 \) plots for H1 and L1 detectors for all injections.}
\label{fig:weighted_avg_pulsar_combined}
\end{figure}

\noindent Figure~\ref{fig:weighted_avg_pulsar_combined} compares the injected and recovered gravitational-wave amplitudes \(h_0\) for all 18 hardware injections, plotted as a function of signal frequency for the H1 (left) and L1 (right) detectors.  In each panel the deep pink “×” marks the true injected \(h_0\), the dark blue square the SFT-weighted average recovered by our narrow-band search, and the adjacent label the pulsar index.  The vertical axis spans nearly three decades of amplitude—from \(\sim10^{-26}\) to \(10^{-23}\), reflecting the
range of both intended SNRs and frequency-dependent detector noise levels. Very good agreement between
intended and recovered amplitudes is visible in these graphs, but a more discriminating global comparison
comes from measuring the fractional differences, namely the differences between recovered and intended amplitudes divided by the intended values.

\vspace{8pt}

\begin{figure}[ht]
\centering
\begin{subfigure}[b]{0.45\linewidth}
    \centering
    \includegraphics[width=\linewidth]{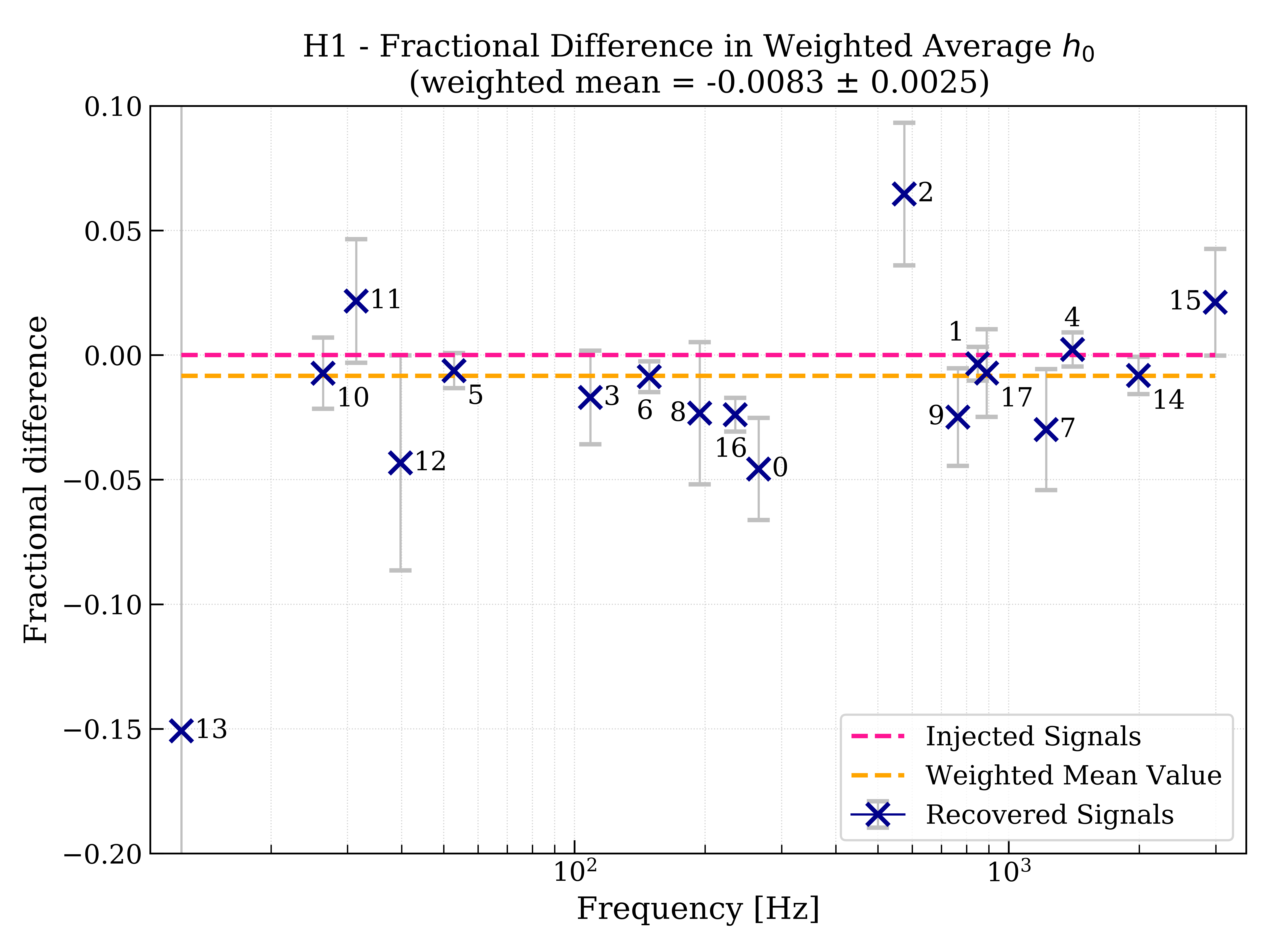}
    \label{fig:H1_relative_amplitudes}
\end{subfigure}
\hfill
\begin{subfigure}[b]{0.45\linewidth}
    \centering
    \includegraphics[width=\linewidth]{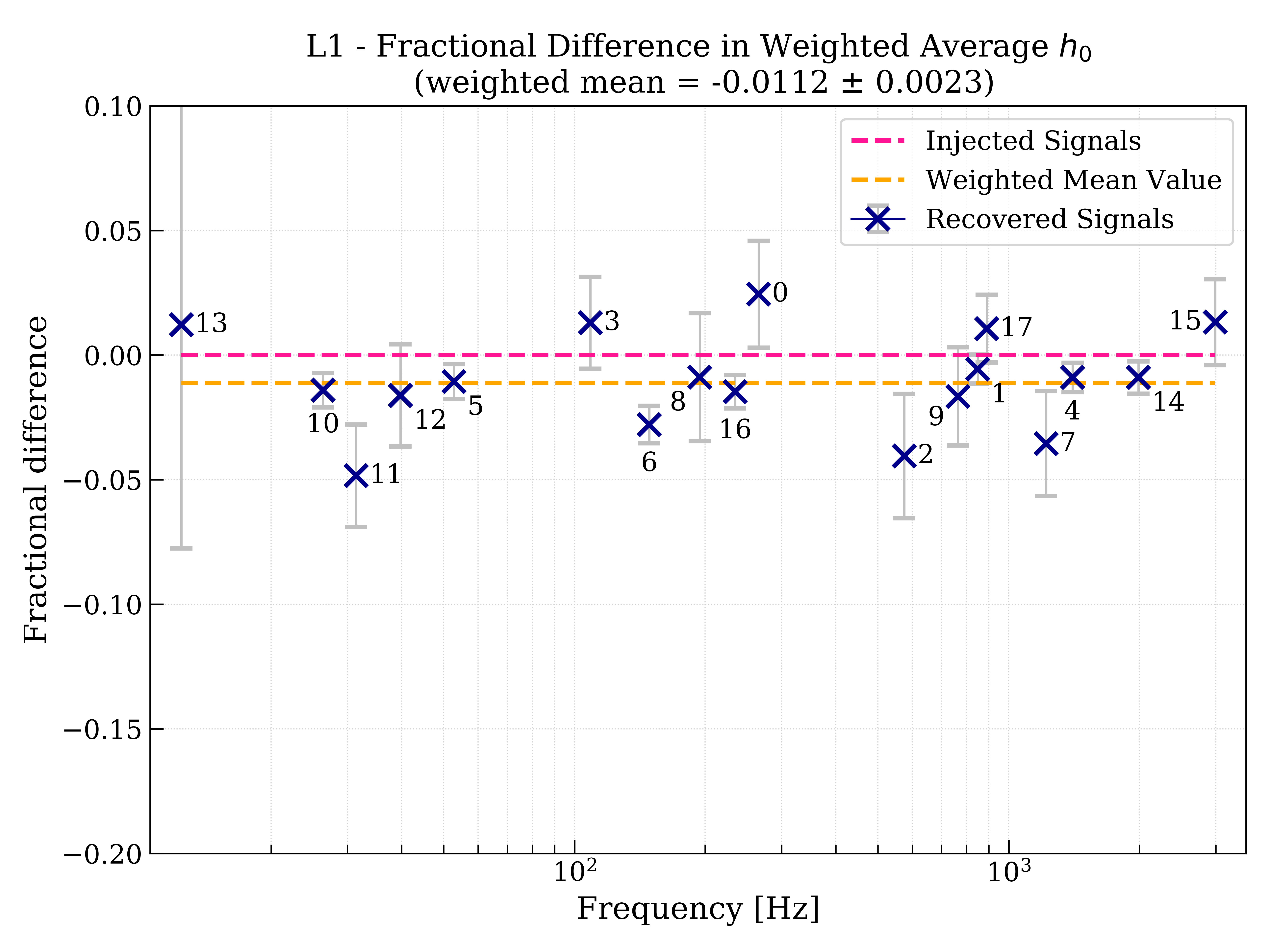}
    \label{fig:L1_relative_amplitudes}
\end{subfigure}

\caption{Relative amplitude deviation and uncertainty plots for H1 and L1 detectors. These plots illustrate the relative amplitude accuracy and precision in the recovered injections.}
\label{fig:relative_amplitudes_combined}
\end{figure}

\noindent These relative amplitudes are shown in Figure~\ref{fig:relative_amplitudes_combined}, with H1 values and uncertainties in the left panel and L1 values in the right panel. For H1, one sees excellent agreement with the ideal value of zero with a weighted mean of -0.0083 $\pm$ 0.0025. For L1, one sees a slight discrepancy of -0.0112 $\pm$ 0.0023, with a corresponding bias toward negative values visible in the graph. This discrepancy is consistent with the estimated uncertainty for the actuation functions determined shortly before the start of O4a running.

\begin{figure}[ht]
\centering
\begin{subfigure}[b]{0.45\linewidth}
    \centering
    \includegraphics[width=\linewidth]{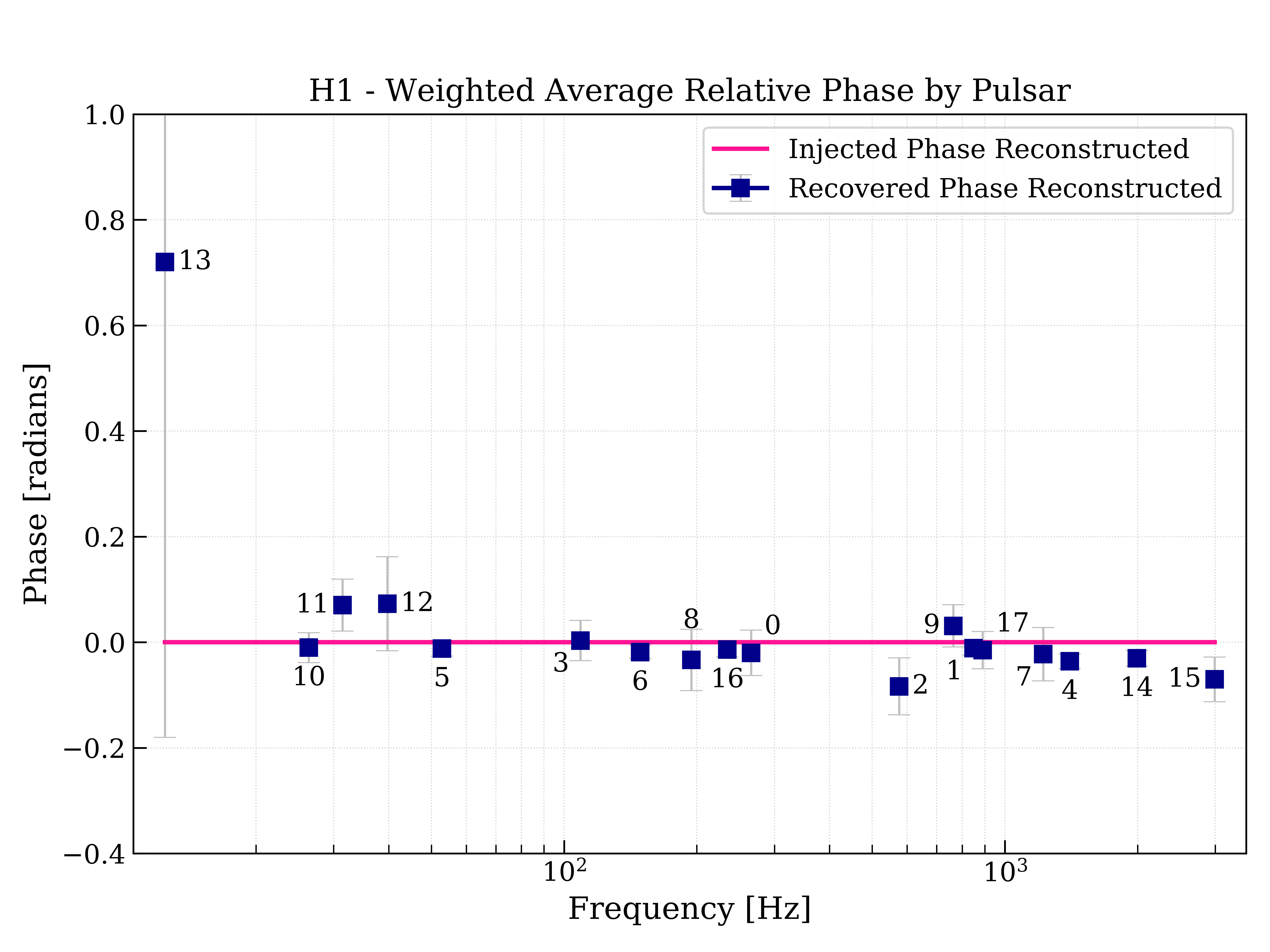}
    \label{fig:H1_weighted_avg_phase_pulsar}
\end{subfigure}
\hfill
\begin{subfigure}[b]{0.45\linewidth}
    \centering
    \includegraphics[width=\linewidth]{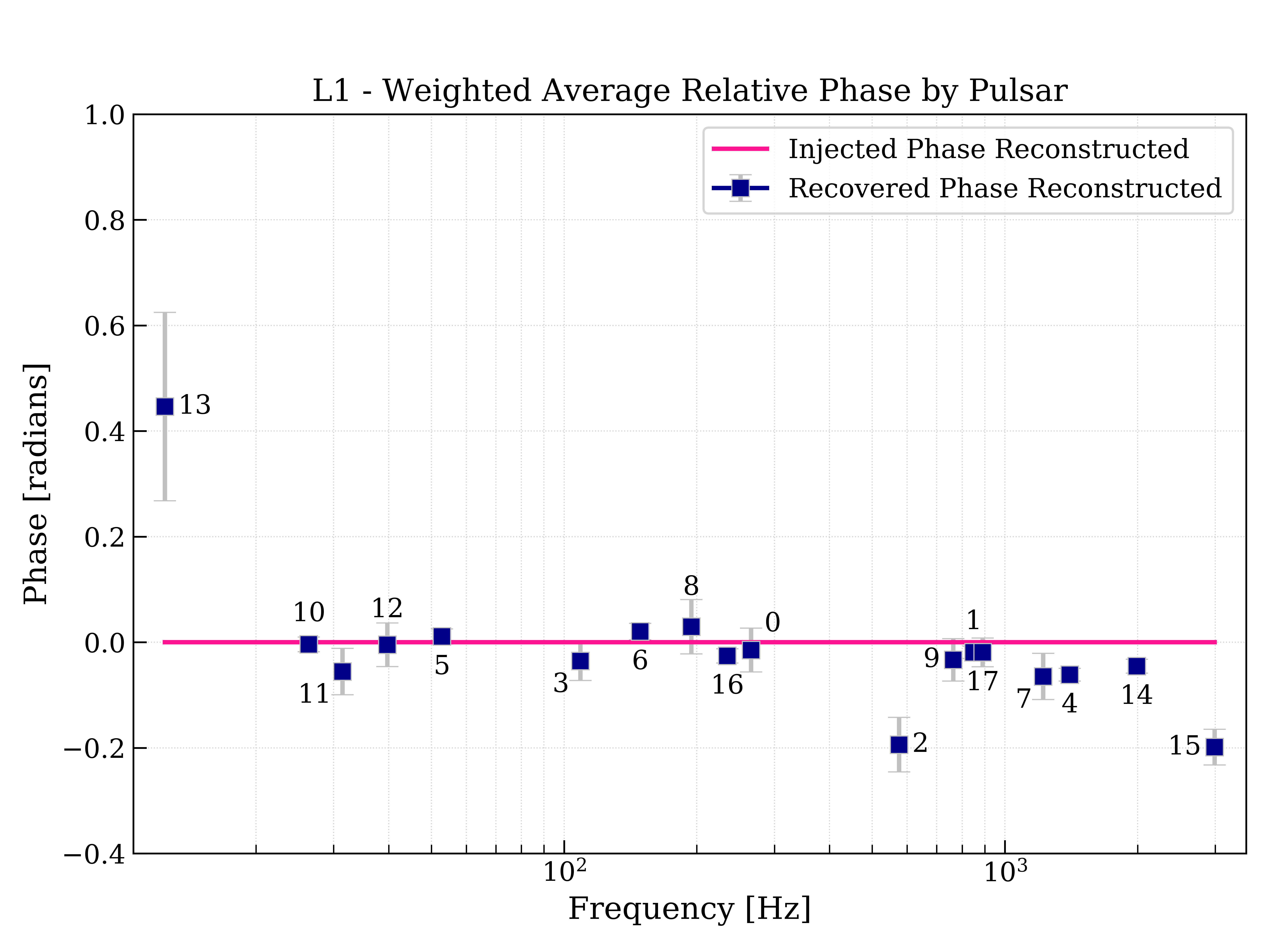}
    \label{fig:L1_weighted_avg_phase_pulsar}
\end{subfigure}

\caption{Weighted average relative phase plots for H1 and L1 detectors. These plots illustrate the phase consistency across pulsars, highlighting any systematic deviations between the detectors.}
\label{fig:weighted_avg_phase_pulsar_combined}
\end{figure}

\begin{figure}[ht]
\centering
\begin{subfigure}[b]{0.45\linewidth}
    \centering
    \includegraphics[width=\linewidth]{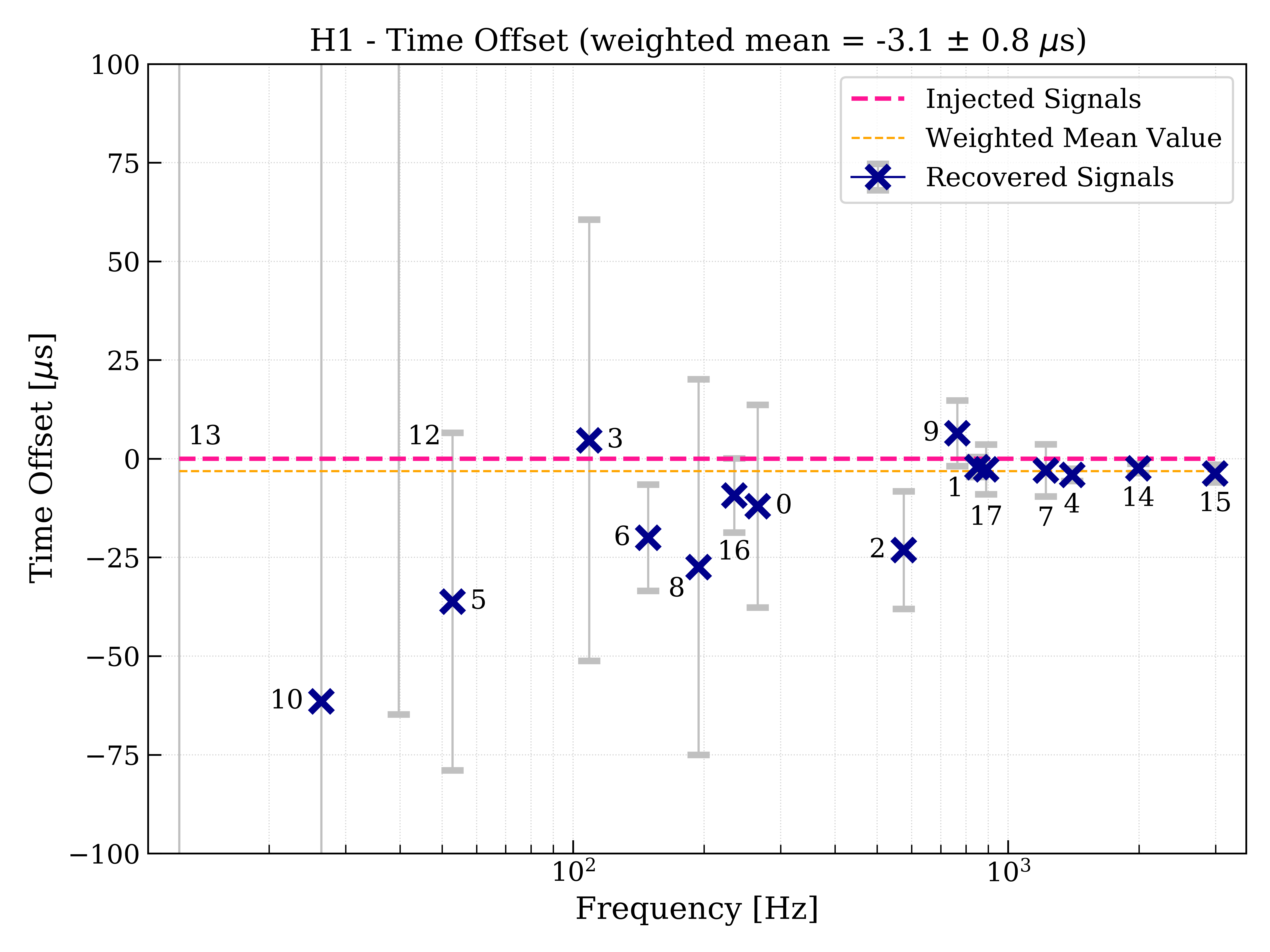}
    \label{fig:H1_time_offset}
\end{subfigure}
\hfill
\begin{subfigure}[b]{0.45\linewidth}
    \centering
    \includegraphics[width=\linewidth]{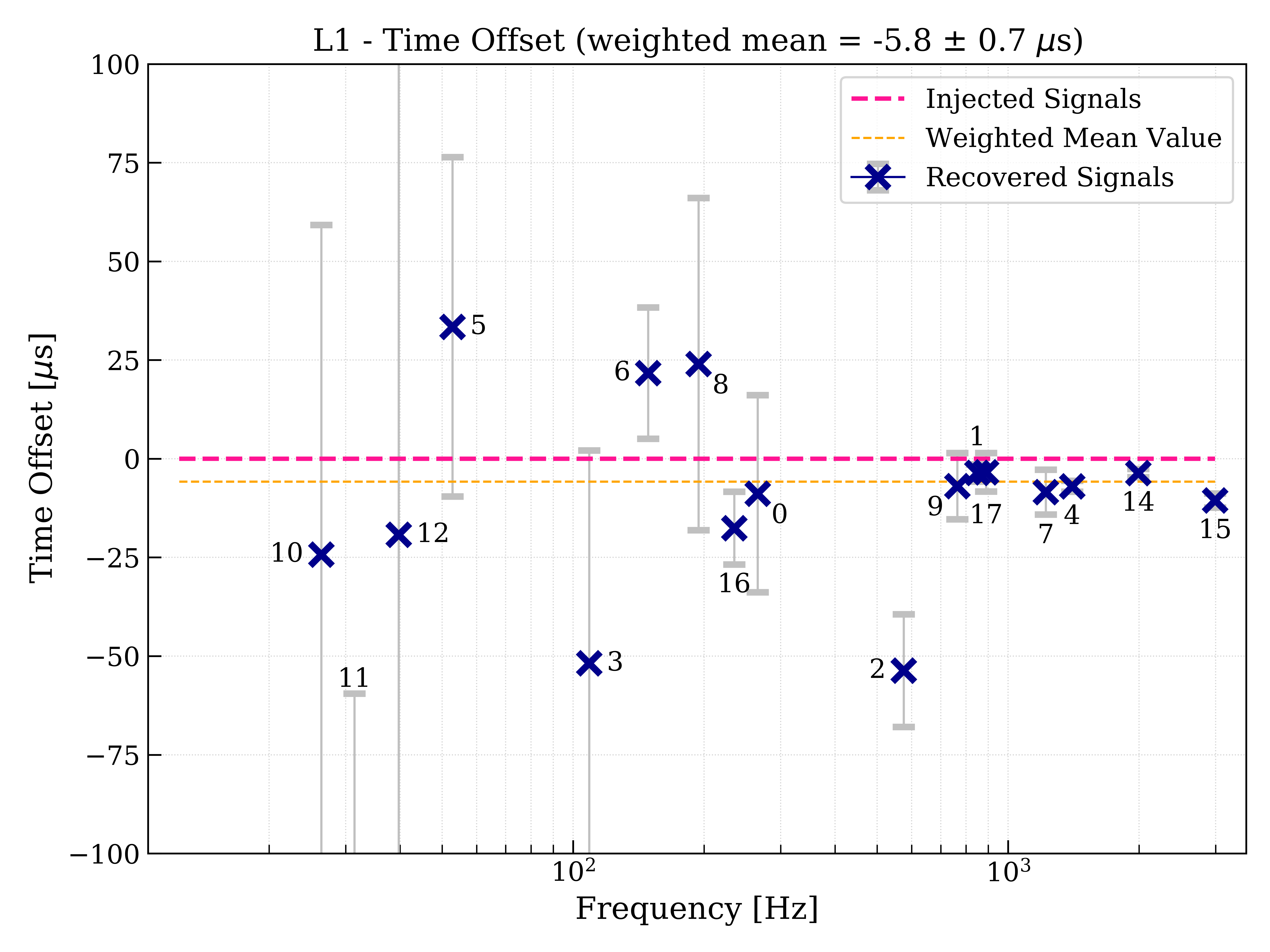}
    \label{fig:L1_time_offset}
\end{subfigure}

\caption{Time offset plots for H1 and L1 detectors. These plots illustrate the relative timing offsets and precisions obtained in the recovered injections.}
\label{fig:time_offset_combined}
\end{figure}

\noindent Similarly, combined phase offset reconstruction graphs provide a validation of the
understanding of timing delays, both in the detector response and in the data stream.
Figure~\ref{fig:weighted_avg_phase_pulsar_combined} shows the reconstructed phases (radians) for
all injections in H1 (left panel) and L1 (right panel). One sees relative phase values near
the ideal of zero. If there were an uncorrected fixed time delay (or advance) in the detector
response, one would expect a phase difference that grows linearly with frequency. To check
for this potential discrepancy, we convert reconstructed relative phase into an equivalent
time offset, as shown ($\mu$s) in Figure~\ref{fig:time_offset_combined}. One sees that
the weighted mean H1 time delay is about 3 $\mu$s, and that for L1 it is about 6 $\mu$s.
Both values lie within the actuation function timing uncertainty estimated prior to the
start of O4a. For reference, the clock cycle for the controls and data acquisition system
is about 61 $\mu$s.

\vspace{8pt}

\begin{figure}[ht]
  \centering
  \includegraphics[width=3.in]{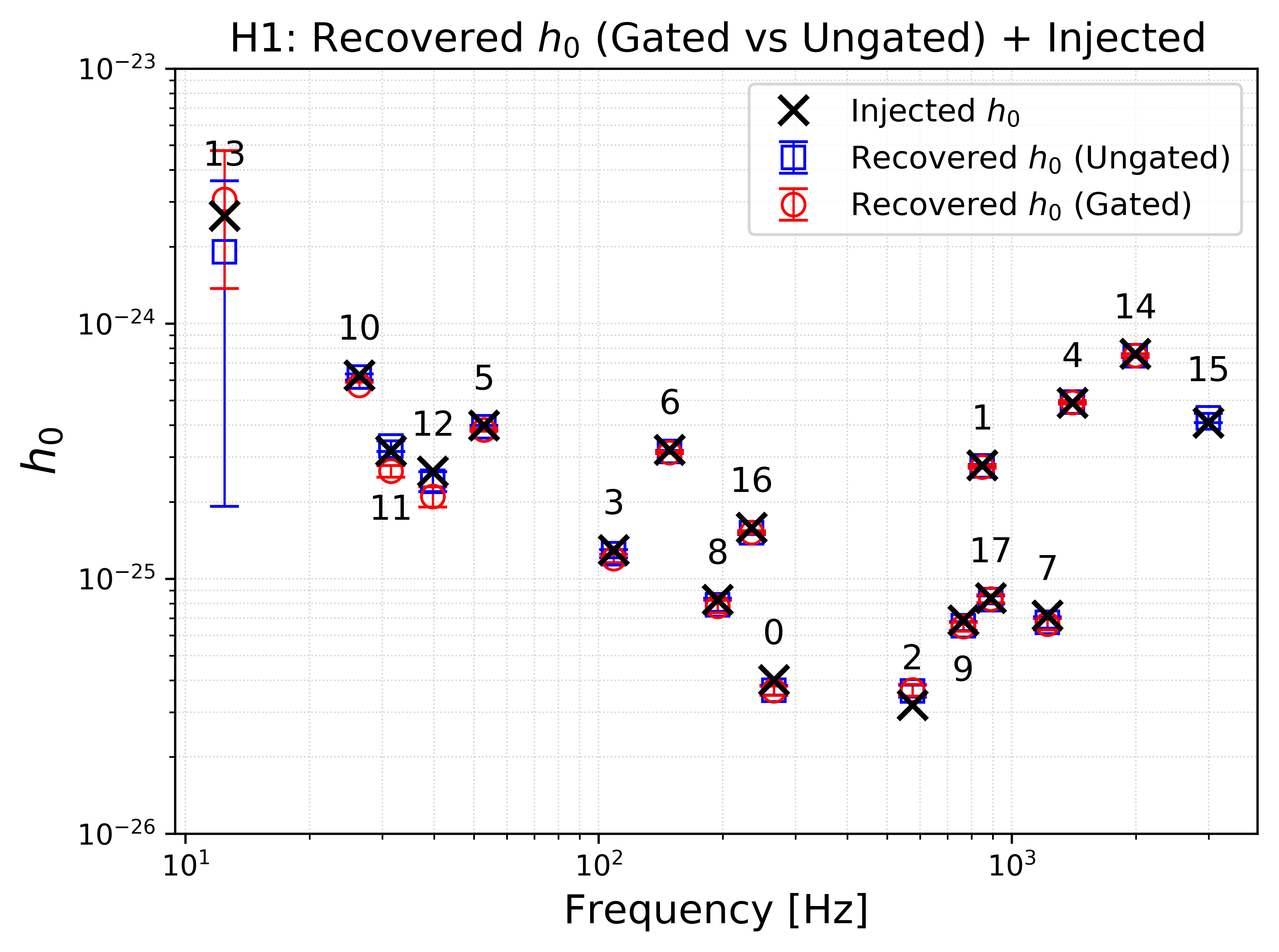}
  \includegraphics[width=3.in]{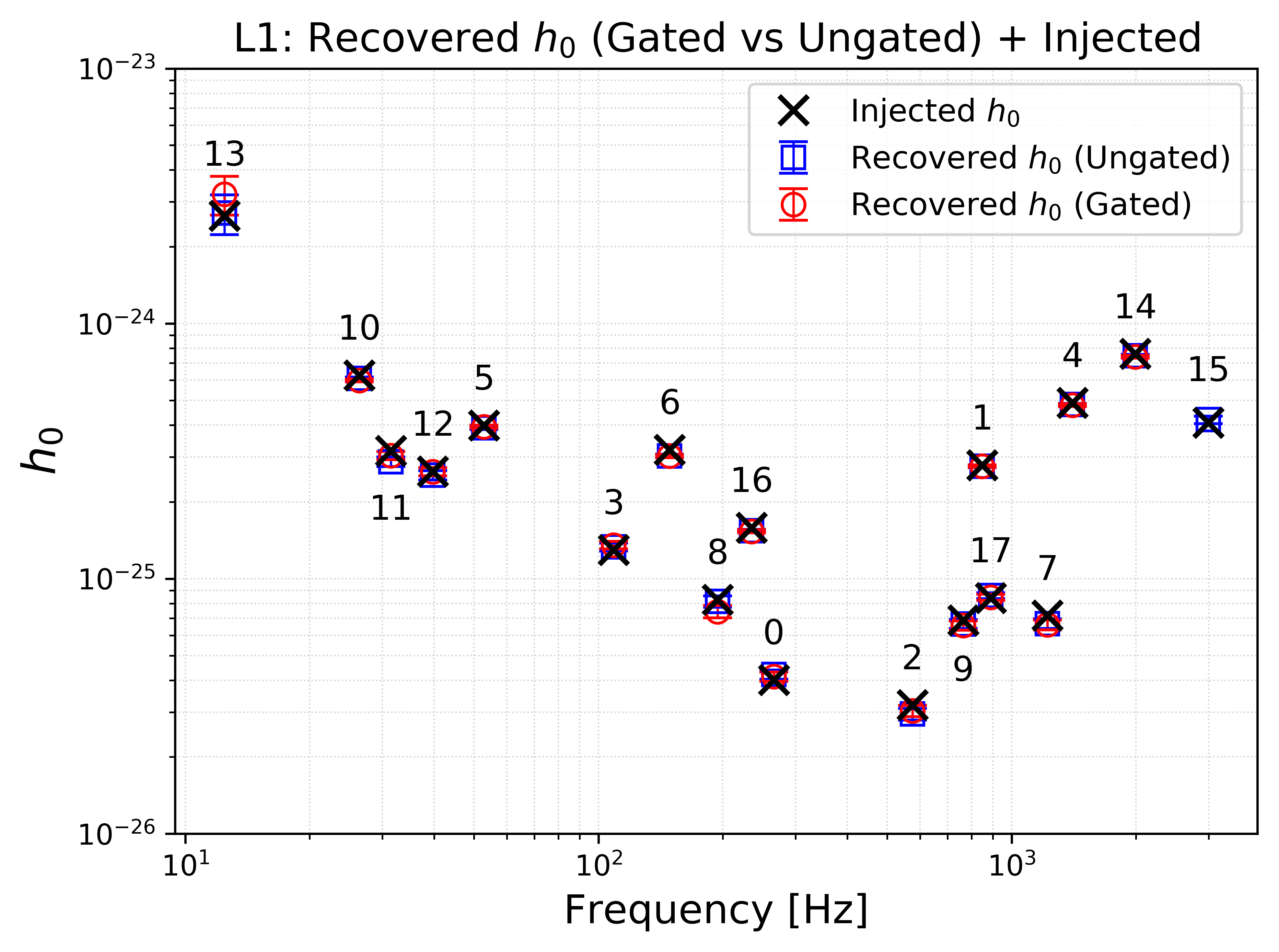}
  \caption{Comparison of signal amplitude reconstruction with and without self-gating in O4a data.
    {\it Left:} H1 reconstruction. 
    {\it Right:} L1 reconstruction. }
  \label{fig:gatedvsungated}
\end{figure}

LIGO data collected during both the O3 and O4 oberving runs have been characterized by
infrequent but extremely loud transient instrumental glitches of uncertain origin, glitches loud enough in their strain power to raise the noise floor below $\sim$500 Hz appreciably even for spectra meausered over time scales of 30 minutes or more. Identification and self-gating of these transients has been carried out for the O3~\cite{zweizig_2021} and O4~\cite{davisetal} data.
The CW hardware injection provide a handy validation of this gating in establishing that SNR for gated data improves relative to that of ungated data at low frequencies and that no statistically significant bias in reconstructed parameters is created by the gating.
Figure~\ref{fig:gatedvsungated} shows, for example, reconstructed $h_0$ values for
H1 and L1 for all 18 injections. As expected, central values change little, but error bars
are somewhat reduced for gated data.

\noindent Collectively, these amplitude and phase offset reconstructions validate the fidelity of
our understanding of the detector response to continuous gravitational waves over as many as tens
of billions of signal cycles in the O4a run period.

\subsection{F-statistic Method}
\label{sec:fstatresults}

\begin{figure}[ht]
  \centering
  \begin{subfigure}[b]{0.85\linewidth}
    \centering
    \includegraphics[width=0.85\linewidth]{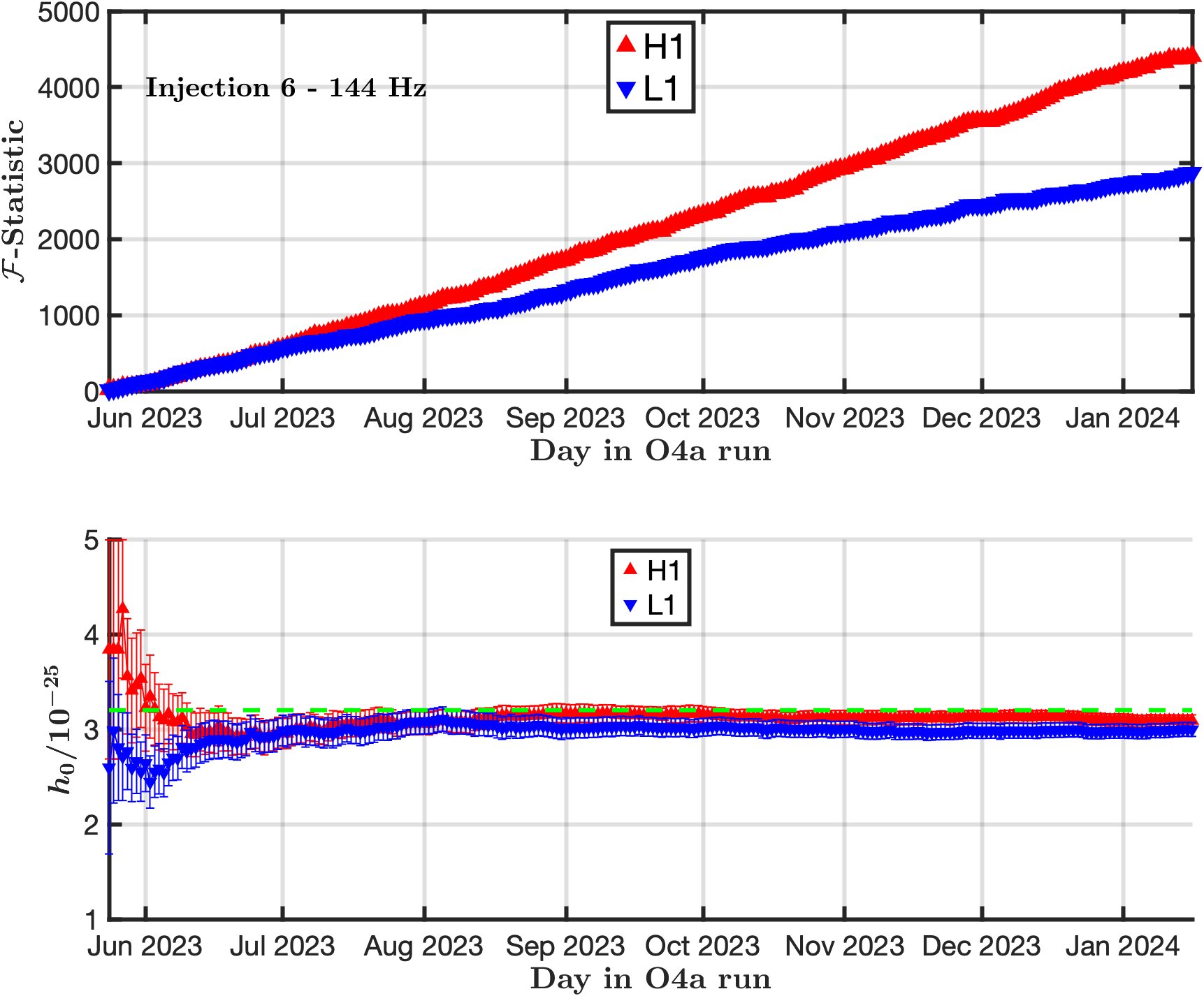}
  \end{subfigure}
  \vskip0.5in
  \begin{subfigure}[b]{0.85\linewidth}
    \centering
    \includegraphics[width=0.85\linewidth]{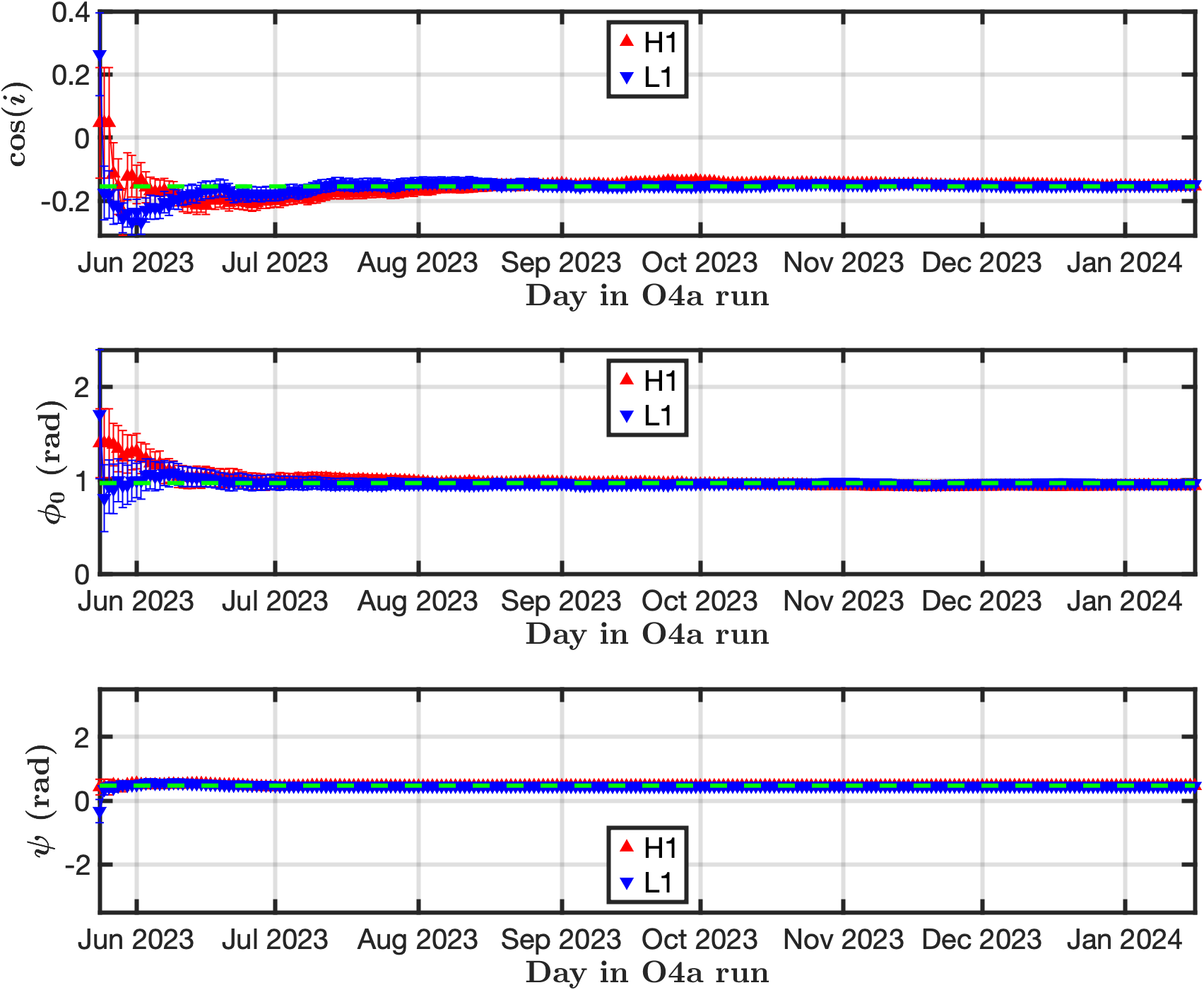}
  \end{subfigure}
  \caption{{\it Top panel:} Cumulative computed \Fstatistic\ values and reconstructed $h_0$ values for Injection 6 ($\sim$144 Hz) in H1 and L1 data \vs\ time in the O4a epoch. {\it Bottom panel:} Corresponding cumulative computed $\cos(\iota)$, $\phi_0$ and $\psi$ values for Injection 6.}
  \label{fig:Inj6}
\end{figure}

Sample results from the \Fstatistic\ monitoring of injections are presented here.
Figure~\ref{fig:Inj6} shows graphs
of cumulative H1 and L1 \fstatistic\ values for Injection 6 ($\sim$144 Hz), one of the loud signals,
through the O4a period, along with estimated $h_0$ values and uncertainties.
One sees a monotonically increasing cumulative \fstatistic\ value, as expected.
For stationary detection noise and stable data collecting livetime, one would expect an approximately linear rise in \fstatistic\ with time.
Downward deviations in slope can arise from poor livetime and/or degraded noise during a given day or days.
The estimated $h_0$ is biased upward early in the run because the maximization over a narrow band can initially sieze upon
an upward statistical fluctuation. Ideally, as the run progresses, the program converges upon the correct frequency bin, as the
estimated amplitude drifts toward the intended value.

\vspace{8pt}

\begin{figure}[ht]
  \centering
  \begin{subfigure}[b]{0.85\linewidth}
  \centering
  \includegraphics[width=0.85\linewidth]{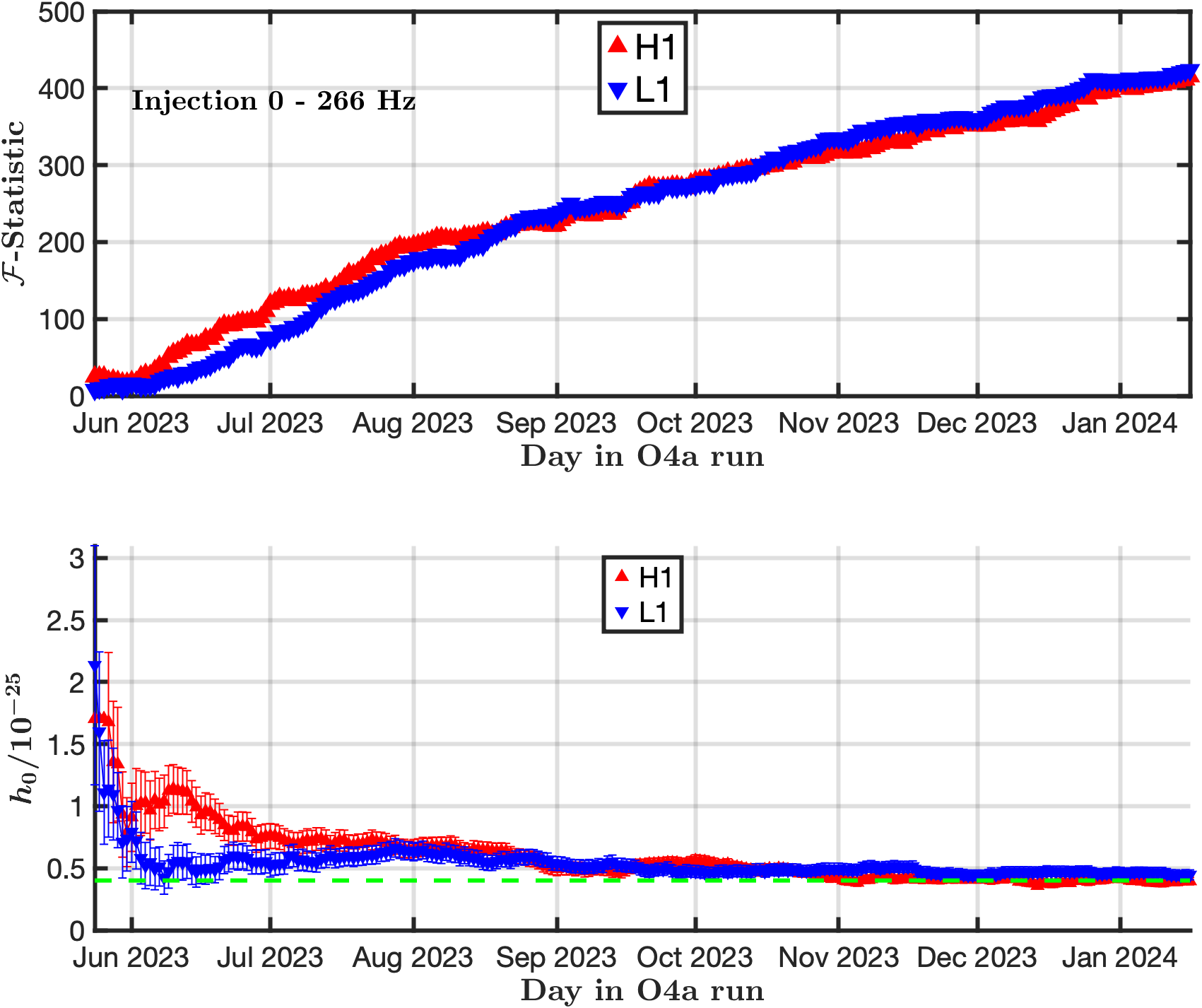}
  \end{subfigure}
  \vskip0.5in
  \begin{subfigure}[b]{0.85\linewidth}
    \centering
    \includegraphics[width=0.85\linewidth]{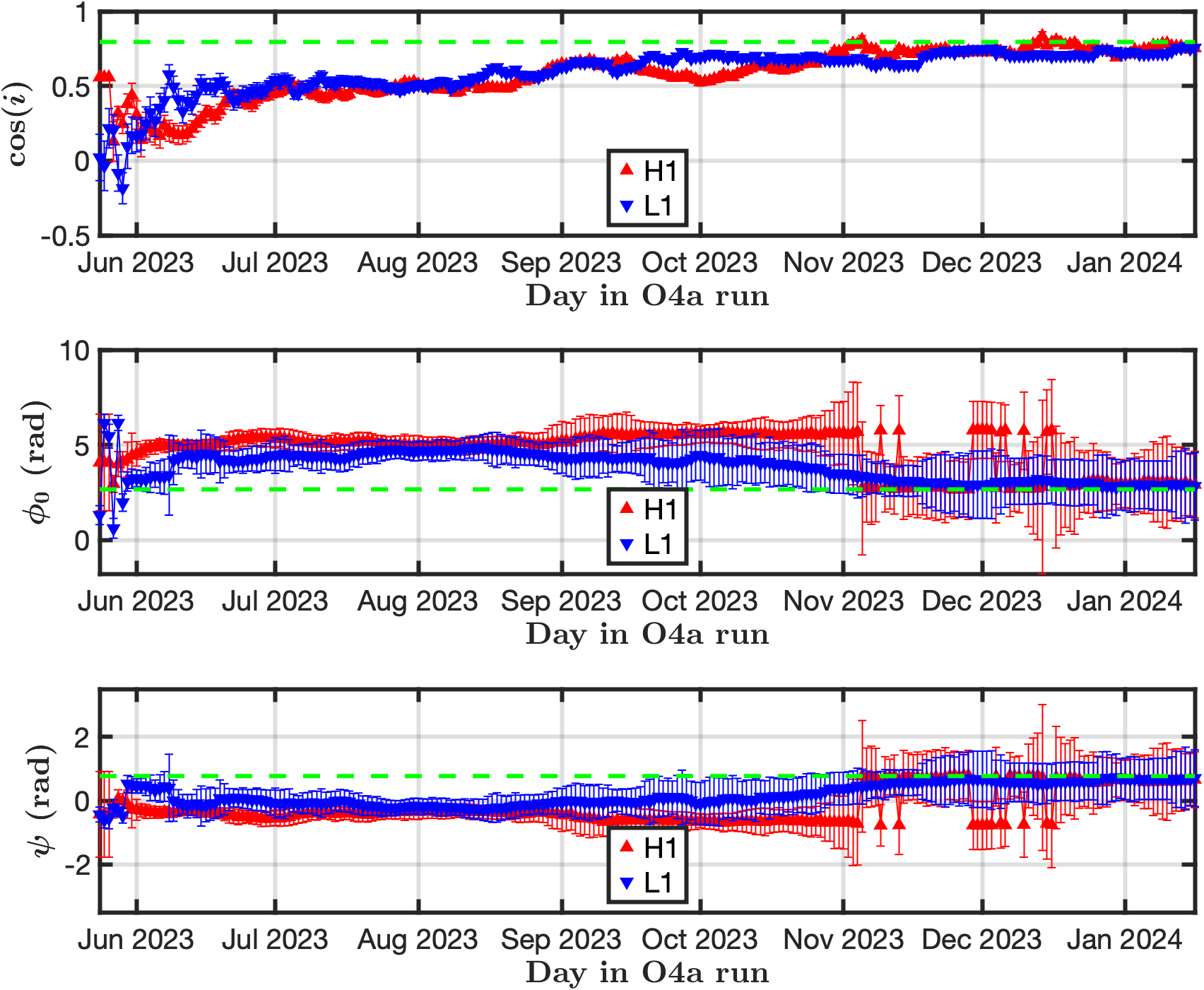}
  \end{subfigure}
  \caption{{\it Top panel:} Cumulative computed \Fstatistic\ values and reconstructed $h_0$ values for Injection 0 ($\sim$266 Hz) in H1 and L1 data \vs\ time in the O4a epoch. {\it Bottom panel:} Corresponding cumulative computed $\cos(\iota)$, $\phi_0$ and $\psi$ values for Injection 0.}
  \label{fig:Inj0}
\end{figure}

\noindent Figure~\ref{fig:Inj6} also shows the cumulative estimated $\cos(\iota)$, $\phi_0$ and $\psi$ for the same Injection 6.
Again, one sees deviations from the intended values that decrease with time, as expected. For this nearly linearly polarized signal
($|\cos(\iota)| \ll 1$), the trends behave as one would naively expect. For nearly circularly polarized signal ($|\cos(\iota)|\approx 1$), on the other hand,
there is a strong anti-correlation between reconstructed $h_0$ and $|\cos(\iota)|$ values, and near degeneracy in the definitions of
the phase offset $\phi_0$ and the polarization angle $\psi$. Figure~\ref{fig:Inj0} shows
corresponding results for an example (Injection 0) with $|\cos(\iota)|$ = 0.795. One sees early in the run
a strong anti-correlation between deviations
of the reconstructed $h_0$ and $\cos(\iota)$ values from their injected values. Similarly, one sees a strong anti-correlation between deviations
of the reconstructed $\phi_0$ and $\psi$ values from their injected values.

\vspace{8pt}

\noindent A rough measure of the detected strength of a signal can be obtained via an {\it effective} strain defined by the quadrature
sums of the magnitudes of the plus- and cross-polarized signals defined in Eqn.~\ref{eqn:polarizations}:
$h_{\rm eff} \equiv \sqrt{h_+^2+h_\times^2}$. From Eqn.~\ref{eqn:polarizations}, one sees that $h_{\rm eff}$ increases monotonically
with $h_0|\cos(\iota)|$, hence the anti-correlation seen in Fig.~\ref{fig:Inj0}.

\vspace{8pt}

\begin{figure}[ht]
\centering
  \begin{subfigure}[b]{0.495\linewidth}
    \centering
    \includegraphics[width=\linewidth]{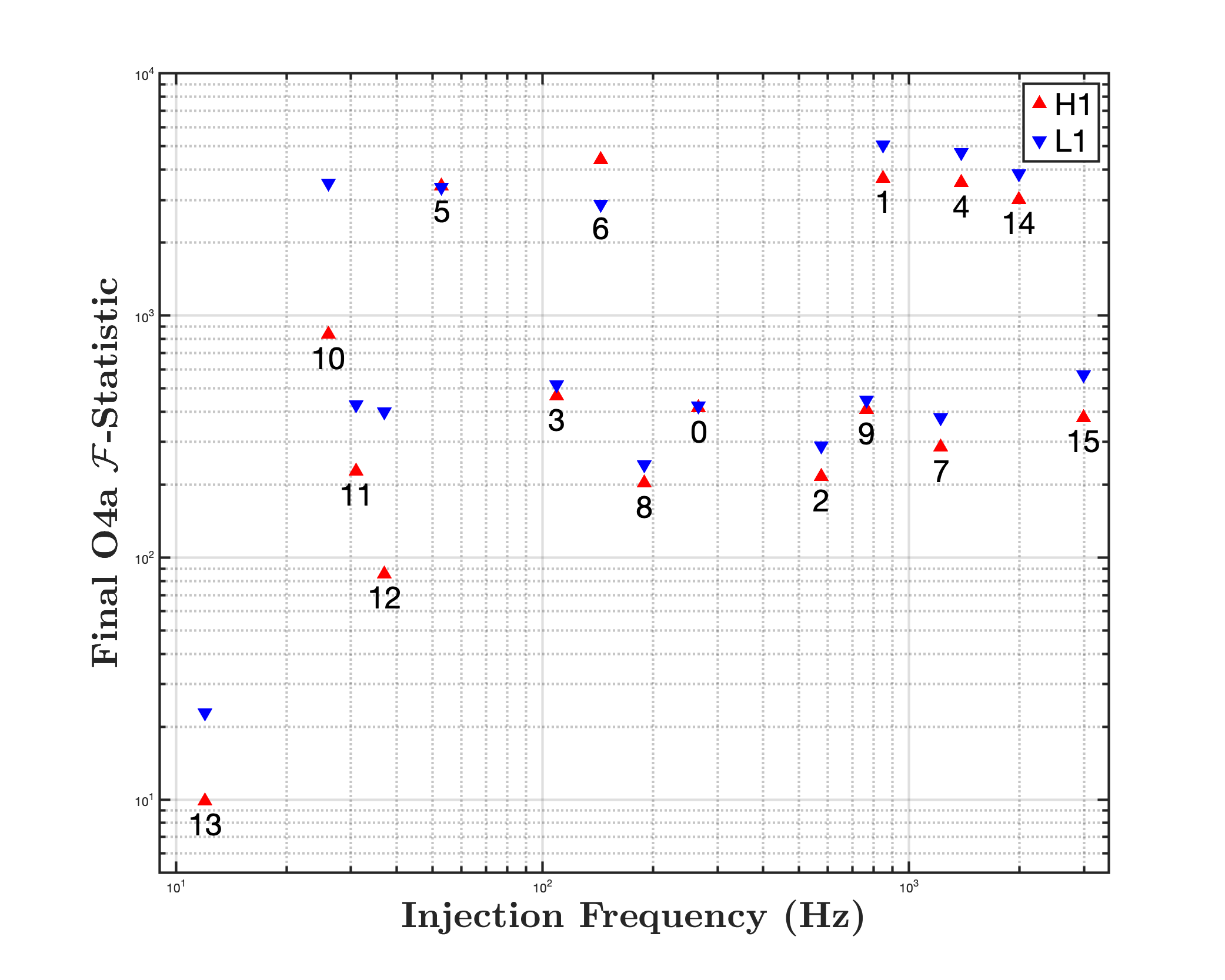}
  \end{subfigure}
  \begin{subfigure}[b]{0.495\linewidth}
    \centering
    \includegraphics[width=\linewidth]{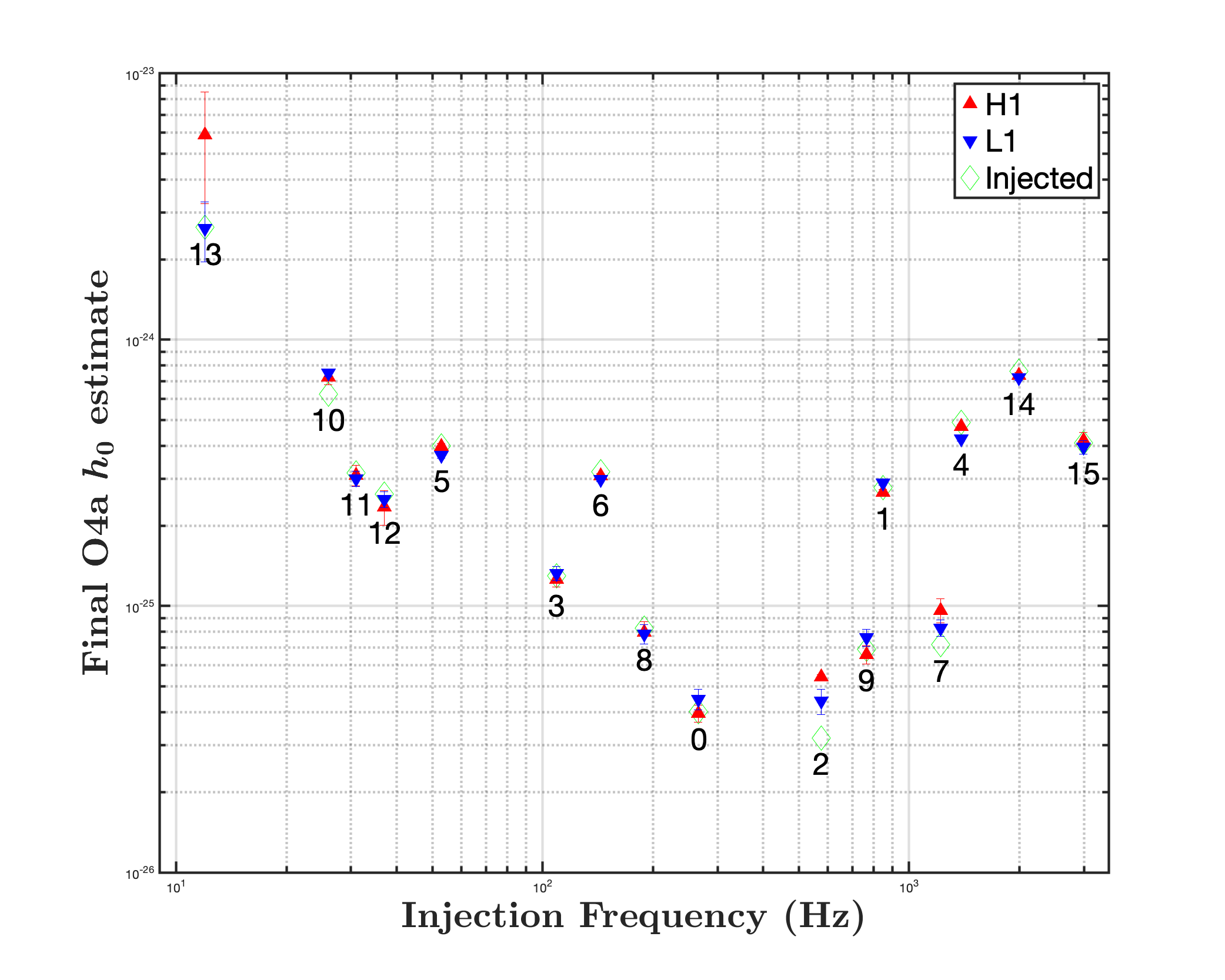}
  \end{subfigure}
  \caption{{\it Left panel:} Final O4a computed \Fstatistic\ values for all isolated injections in H1 and L1 data.
  {\it Right panel:} Corresponding inal O4a reconstructed $h_0$ values for all isolated injections.}
  \label{fig:Fstat_h0_Summary}
\end{figure}

\noindent Figure~\ref{fig:Fstat_h0_Summary} shows the \fstatistic\ values recovered for all of the isolated-star injections
for both H1 and L1 by the end of O4a. The values obtained for the two detectors differ because of different noise
levels and integrated livetimes. For most injections, the recovered L1 values are higher. The loudest six injections
used for daily monitoring of unintended calibration changes are apparent, especially
for L1. Figure~\ref{fig:Fstat_h0_Summary} also shows the $h_0$ values recovered for the same injections in both detectors,
along with the intended values. One sees good agreement for most injections (albeit with large uncertainties for Injection 13
12.34 Hz which is in a band marked by highly non-Gaussian artifacts. The region below 20 Hz has not traditionally been used for CW searches.

\begin{figure}[ht]
  \centering
  \begin{subfigure}[b]{0.495\linewidth}
    \centering
    \includegraphics[width=\linewidth]{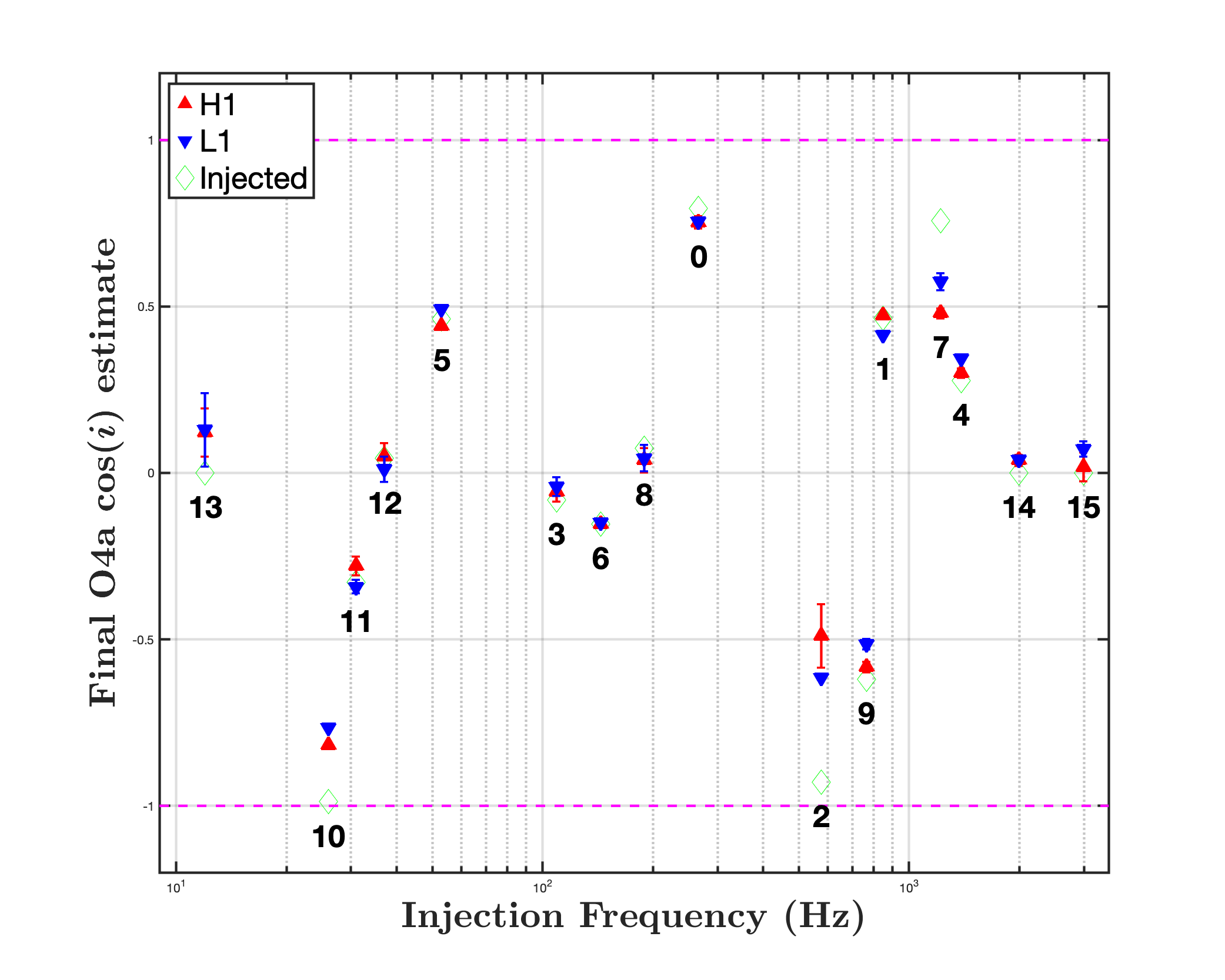}
  \end{subfigure}
  \begin{subfigure}[b]{0.495\linewidth}
    \centering
    \includegraphics[width=\linewidth]{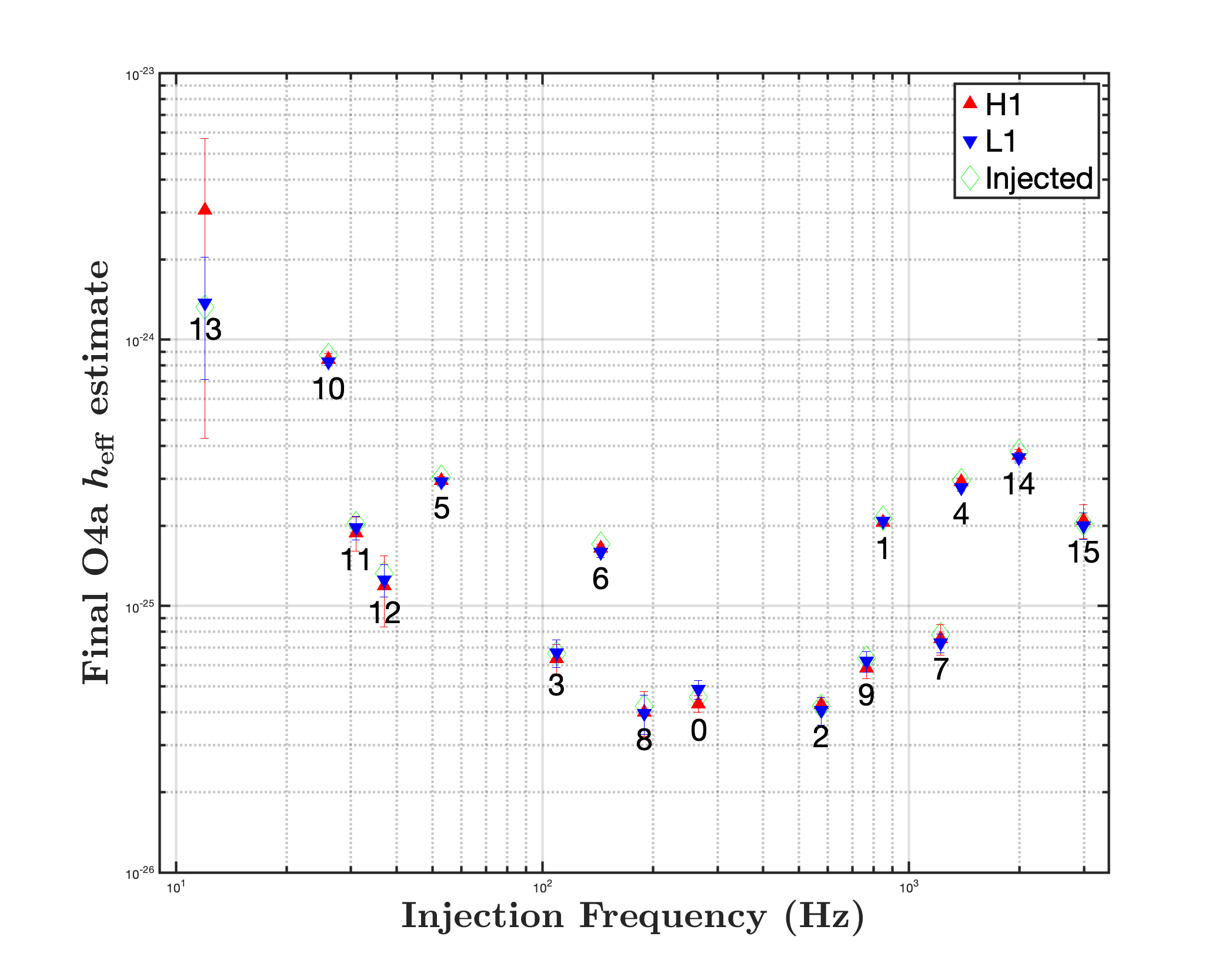}
  \end{subfigure}
  \caption{{\it Left panel:} Final O4a reconstructed $\cos(\iota)$ values for all isolated injections in H1 and L1 data. {\it Right panel:} Corresponding final O4a reconstructed $h_{\rm eff}$ values for all isolated injections. }
  \label{fig:cosi_heff_Summary}
\end{figure}

\noindent Notable discrepanices, however, are seen for Injections 2, 7 and 10. These discrepancies in reconstructed $h_0$ are reflected
in Fig.~\ref{fig:cosi_heff_Summary} which shows corresponding reconstructed $\cos(\iota)$ values. Combining $h_0$ and $\cos(\iota)$
into effective strain $h_{\rm eff}$ yields much better agreement, as expected, which is shown in Fig.~\ref{fig:cosi_heff_Summary},
where excellent consistency is seen for every injection. Nonetheless, the disentanglement of the four source parameters reconstructed
in single-detector \Fstatistic\ analysis highlights the value of the templated monitoring method for clean validation of amplitude and
phase response. 

\subsection{Bayesian Method}

\begin{figure}[ht]
    \centering
    \includegraphics[width=0.95\linewidth]{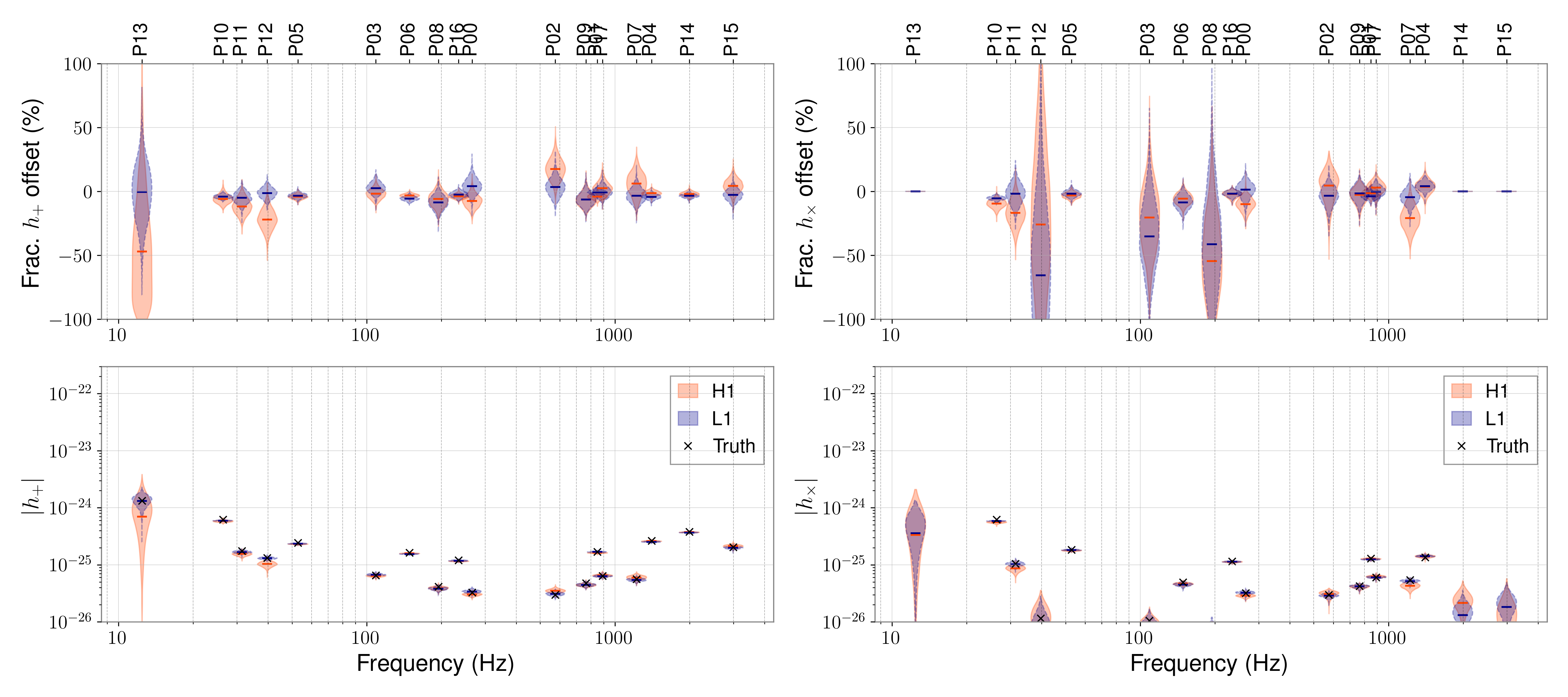}
    \caption{Violin plots of the posterior probability distributions of the inferred $h_+$ and $h{\times}$
    amplitudes for each hardware injection as recovered using the Bayesian method as a function of injection signal frequency. The top panels show the
    posteriors as a fractional offset from the expected injection values, while the bottom panel shows the
    posteriors themselves. The recovered posteriors for H1 and L1 are shown in blue and orange, respectively.}
    \label{fig:bayesian_amplitudes}
\end{figure}

The hardware injections for both H1 and L1 were recovered using data between 2023-05-24 15:00:00
and 2024-01-16 16:00:00. For each injection and detector, the data was heterodyned as described in Section~\ref{sec:bayesian_method} and the posterior probability distributions for
$\{h_0, \cos{\iota}, \psi, \Phi_0\}$ were estimated. Of particular interest are the recovered amplitude and
initial phase. In Figure~\ref{fig:bayesian_amplitudes}, the recovered posteriors on $h_0$ and $\cos{\iota}$ are recast as posteriors on $h_+$ and $h_{\times}$, as in Eqn.~\ref{eqn:polarizations}, and the distributions are shown as violin
plots of the fractional offset from the expected values and the true absolute values. The narrower the distribution, the more precisely the amplitude could be recovered. For the fractional offset, the closer to zero means that
the recovered injection amplitude more closely matched the expected value. We see no specific trend in amplitude offsets
as a function of frequency and the posteriors are broadly consistent with the expected amplitudes.

\begin{figure}[ht]
    \centering
    \includegraphics[width=0.65\linewidth]{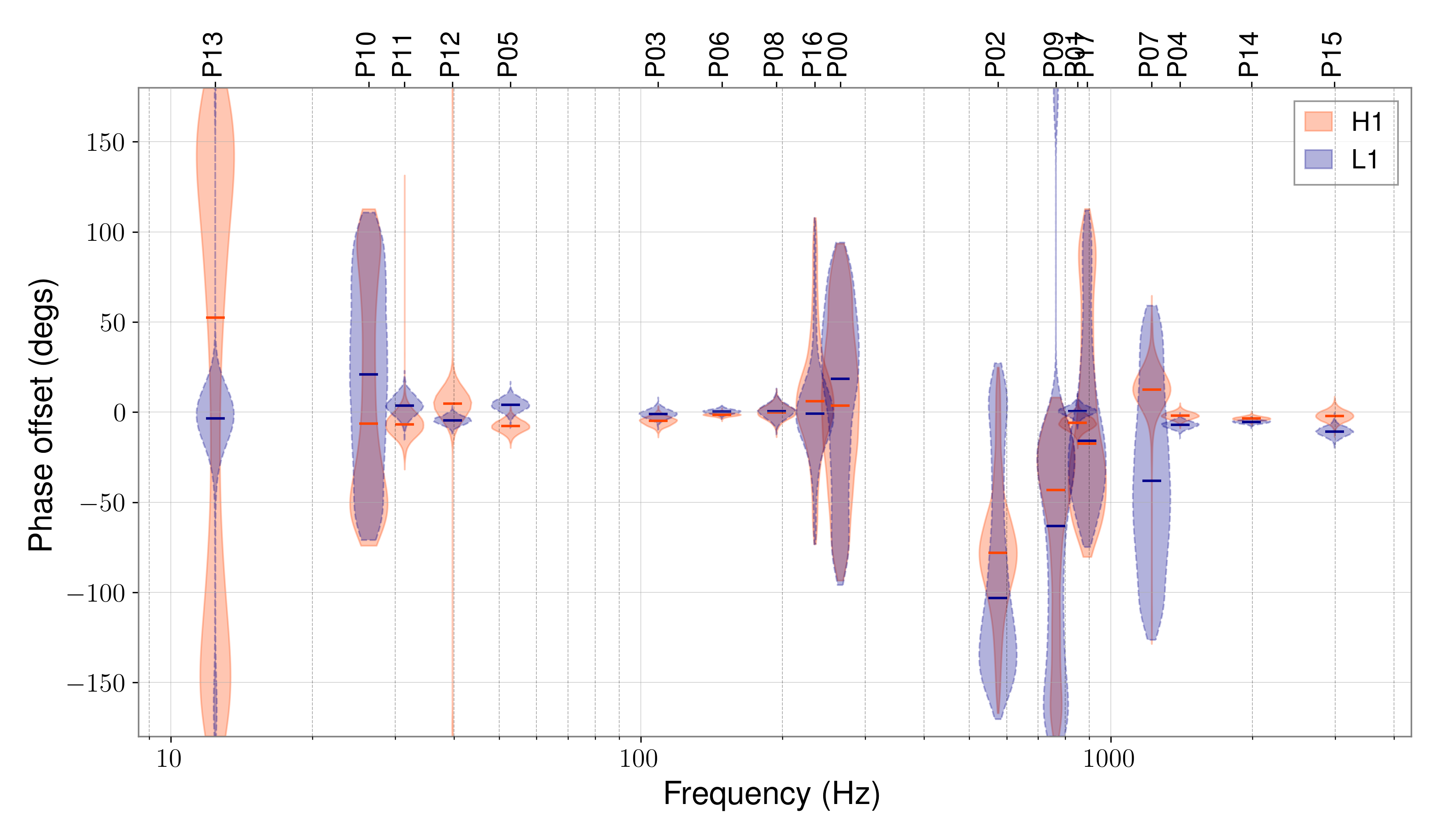}
    \caption{Violin plots of the posterior probability distributions, for each injection, of the inferred offset of the recovered initial phases form the expected injection value as a function of injection signal frequency. The recovered posteriors for H1 and L1 are shown in blue and orange, respectively.}
    \label{fig:bayesian_phases}
\end{figure}

\noindent In Figure~\ref{fig:bayesian_phases}, we similarly see violin plots of posterior distributions
representing the difference between the recovered $\Phi_0$ distribution and the expected injection value. There
is no trend in the offset as a function of injection signal frequency and the posteriors are broadly consistent with
the injected value. The broadest posteriors generally reflect the high degree of correlation between the $\Phi_0$
and $\psi$ which occurs as the signal becomes closer to circularly polarized, i.e., as $h_{\times}$ comes to dominate
the amplitude.

\section{Conclusion}

In summary, this paper presents three techniques used in the real-time monitoring and analysis of hardware injections in LIGO data, highlighting their role in validation of searches for astrophysical sources of
continuous gravitational waves.

\vspace{8pt}

\noindent The first, templated technique applies all of the expected frequency
and amplitude modulations, in order to achieve the best precision on absolute amplitude and phase offset.
The second technique, based on the long-established \Fstatistic, allows for unknown stellar orientation
angles and hence unknown amplitude modulations. The third, Bayesian inference technique also allows
for unknown orientations and presents posterior distribution functions. The complementary approaches
and the mostly independent code bases give further confidence in the validation.

\vspace{8pt}

\noindent As we move closer to the detector sensitivity necessary for the
first-ever detection of an astrophysical CW source, it is critical to 
verify our search and signal reconstruction techniques, including understanding detector
response well enough to phase-track a source over billions of signal cycles. 

\section{Acknowledgments}

We gratefully acknowledge useful discussions and long collaboration with current
and former colleagues in the LIGO-Virgo-KAGRA continuous waves working group and the LIGO calibration working group.
We thank, in particular, David Barker, Joseph Betzweiser, Evan Goetz, Michael Thomas
and Keith Thorne for ensuring reliable, nearly 24/7 LIGO hardware injections during
the O4 run. We also thank Evan Goetz for helpful suggestions and comments on this manuscript.
This work was supported in part by National Science
Foundation Awards PHY-2110181 and PHY-2408883.

\vspace{8pt}

\noindent This material is based upon work supported by NSF’s LIGO Laboratory
which is a major facility fully funded by the National Science Foundation.
LIGO was constructed and is operated by the California Institute of Technology and Massachusetts
Institute of Technology with funding from the U.S. National Science Foundation under grant PHY-
0757058.
The authors also gratefully acknowledge the support of the Science
and Technology Facilities Council (STFC) of the United Kingdom,
the Max-Planck-Society (MPS), and the
State of Niedersachsen/Germany for support of the construction of Advanced LIGO.
The authors are grateful for computational resources provided by the
LIGO Laboratory and supported by
National Science Foundation Grants PHY-0757058 and PHY-0823459.

\printbibliography

\end{document}